\documentclass[reprint,
                amsmath,
                amssymb,
                superscriptaddress,
                onecolumn]{revtex4-2}

\usepackage{graphicx}
\usepackage{bm}
\usepackage{xcolor}
\usepackage{mathtools}
\usepackage{physics}
\usepackage{tabularx}
\usepackage{setspace}
\usepackage{hyperref}
\hypersetup{
    colorlinks=true,
    linkcolor=blue,
    citecolor=blue,
    filecolor=magenta,      
    urlcolor=blue,
    pdftitle={Overleaf Example},
    pdfpagemode=FullScreen,
    }

\usepackage{multibib}
\newcites{SI}{Supp Info}

\usepackage{fancyhdr}

\fancypagestyle{head}{%
    \lhead{Non-peer reviewed preprint}
    \rhead{Published version in Nature Geoscience: \url{doi.org/10.1038/s41561-025-01672-w}}
}%
\begin{document}

\thispagestyle{head}

\title[Article Title]{Emergence of wind ripples controlled by mechanics of grain–bed impacts}

\author{C. W. Lester}
\email[Corresponding author contact: ]{conner.lester@duke.edu}
\affiliation{Earth and Climate Sciences, Duke University, Durham, NC 27708, USA}
\author{A. B. Murray}
\affiliation{Earth and Climate Sciences, Duke University, Durham, NC 27708, USA}
\author{Orencio Duran}
\affiliation{Department of Ocean Engineering, Texas A\&M University, College Station, TX 77843-3136, USA}
\author{B. Andreotti}
\affiliation{Laboratoire de Physique de l'Ecole Normale Sup\'erieure, PSL, Sorbonne Universit\'e, Universit\'e Paris Cit\'e, CNRS, 24 rue Lhomond 75005 Paris,  France}
\author{P. Claudin}
\affiliation{Physique et M\'ecanique des Milieux H\'et\'erog\`enes, PMMH UMR 7636 CNRS, ESPCI Paris, PSL Research University, Sorbonne Universit\'e, Universit\'e Paris Cit\'e, 7 quai St Bernard, 75005 Paris, France}

\maketitle

\section*{Abstract}
Periodic sediment patterns have been observed on Earth in riverbeds and sand and snow deserts, but also in other planetary environments. One of the most ubiquitous patterns, familiar wind or ‘impact’ ripples, adorns sand beaches and arid regions on Earth. The observation of aeolian impact ripples on Mars the same size as their terrestrial counterparts despite a thinner atmosphere raises questions about their formation. Here we show in a numerical simulation that the emergent wavelength of impact ripples is controlled by the mechanics of grain–bed impacts and not the characteristic trajectories of grains above the bed. We find that the distribution of grain trajectories in transport is essentially scale-free, invoking the proximity of a critical point and precluding a transport-related length scale that selects ripple wavelengths. By contrast, when a grain strikes the bed, the process leading to grain ejections introduces a collective granular length scale that determines the scale of the ripples. We propose a theoretical model that predicts a relatively constant ripple size for most planetary conditions. In addition, our model predicts that for high-density atmospheres, such as on Venus, or for sufficiently large sand grains on Earth, impact ripples propagate upwind. Although wind-tunnel and field experiments are needed to confirm the existence of such ‘antiripples’, we suggest that our quantitative model of wind-blown sediment transport may be used to deduce geological and environmental conditions on other planets from the sizes and propagation speeds of impact ripples.

\section*{This is a preprint; it has not been peer reviewed by a journal.}
 A version of this preprint was published at Nature Geoscience on April 8th, 2025. See \url{https://doi.org/10.1038/s41561-025-01672-w}.



\begin{figure*}[t]
\centering
 \includegraphics[width=\textwidth]{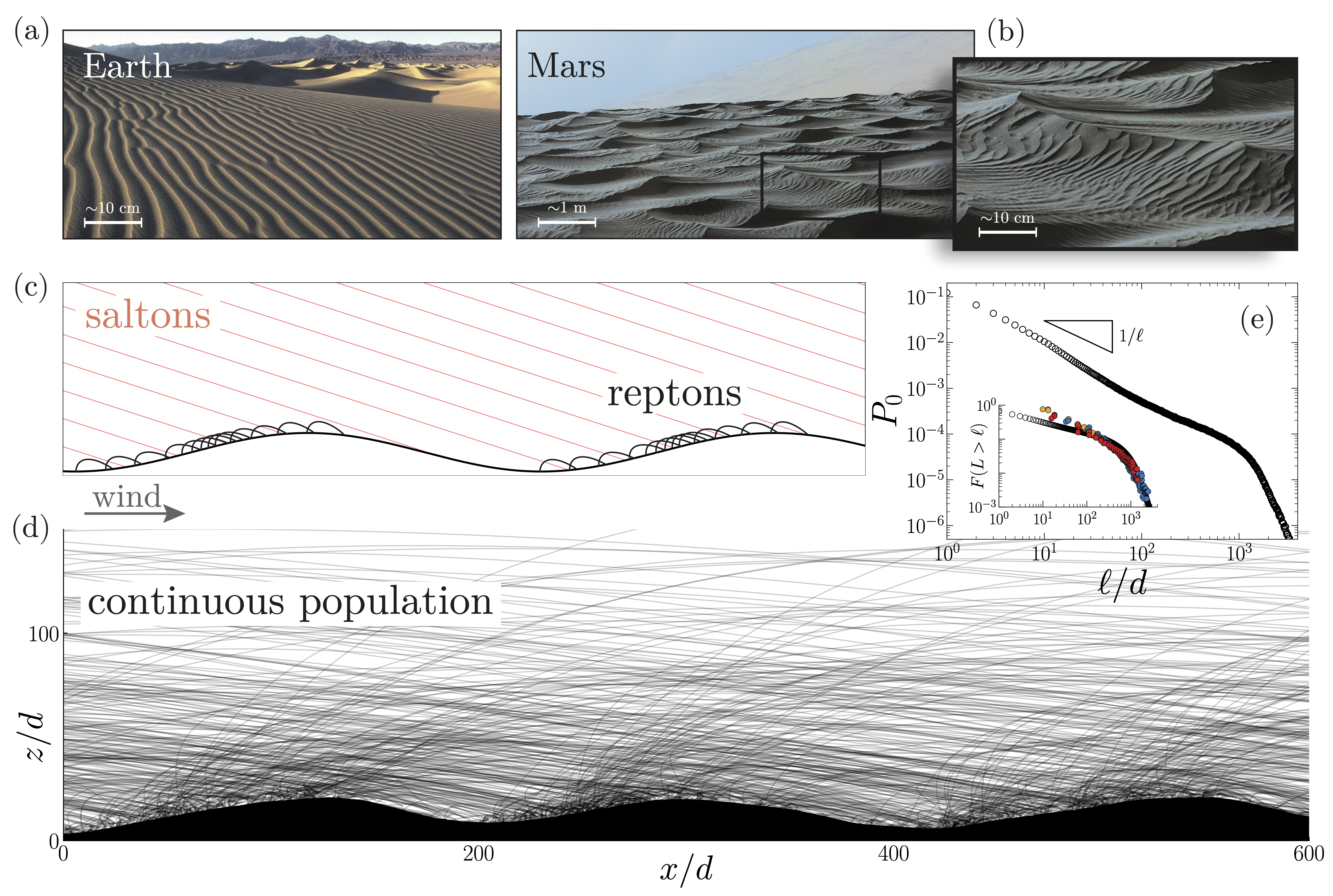}
\caption{Wind ripple patterns have been explained by a two population model, but transport is better described by a continuous scale-free population. (a) Photos of centimeter scale impact ripples on Earth (photo by Marli Miller from Death Valley, CA, USA: geologypics.com) and (b) meter scale ``megaripples'' with superimposed centimeter scale impact ripples (zoomed panel) on Mars (photo from NASA's Curiosity Rover). The ``megaripple'' bedforms on Mars are dynamically distinct from the superimposed impact ripples---as the dunes on Earth are dynamically distinct from the super imposed impact ripples in (a) \cite{lapotre2016large,duran2019unified}. (We note that impact ripple sizes in (a) are mature and around a factor of two or so larger than the initial wavelengths modeled in this paper \cite{andreotti2006aeolian}. And ripples in (b) seem to show a range of sizes, the smallest being the incipient wavelengths addressed here.) (c) The widely used model for sediment transport and ripple formation that separates transport into two distinct populations, saltons and reptons. (d) DEM trajectories over a rippled surface and (e) hoplength distribution $P_0(\ell)$ versus hoplengths $\ell$ rescaled by the grain size $d$ (data from the DEM for Earth like conditions; $\langle\ell\rangle\simeq 100d$). DEM simulations (panels d,e) show that the population of trajectories forms one continuous distribution---much of it scale free as $P_0\sim 1/\ell$---rather than distinct, scale-separated populations as would be observed in the salton-repton model (panel c). Inset in (e) shows the exceedance probability of hoplengths $F(L>\ell)$ compared with wind tunnel experiments of \cite{ho2014aeolian}---for all wind speed runs for $230\ \mu$m (blue) and $630\ \mu$m (red) grain sizes--and wind-tunnel experiments of \cite{rasmussen2015laboratory} (red) for $320\ \mu$m for all wind speed tested.}
\label{fig1}
\end{figure*}

Many approximately periodic sedimentary patterns result from the interaction between bedform and flow, via modulation of transport \cite{charru2013sand,duran2019unified}. There is already a long scientific history of relating bedforms observed in different environments to physical parameters such as fluid density, viscosity and velocity, and grain density and size \cite{bagnold2012physics,charru2013sand,lapotre2016large,duran2019unified}. The search for analogs between different planets involves correctly identifying the dynamical mechanisms. For example, the meter scale dunes on Mars (Fig. \ref{fig1}b) are not the counterparts of centimeter scale wind ripples on Earth (Fig. \ref{fig1}a), but are more similar to ripples in water and sand dunes more broadly \cite{lapotre2016large}. While dunes, water ripples and meter-scale Martian bedforms arise because fluid forces on the bed vary across the bedform, wind ripples, or `impact ripples', form under conditions in which grains at the bed surface are set in motion by high energy impacts of other grains in transport---as opposed to directly impelled by the fluid. On both Earth and Mars, impact ripples are centimeter scale (Fig. \ref{fig1}a,b). However, experimental results suggest that under some conditions wind modulation in response to bed undulations can lead to the coevolution of impact \emph{and} hydrodynamic ripples of similar scale \cite{yizhaq2024coevolving}. To disentangle these effects we choose to analyze impact-ripple formation in the absence of wind modulations. 

The current understanding of impact ripples often attributes their size (wavelength) to the characteristic hoplength of sand grains in transport \cite{bagnold2012physics,anderson1987theoretical,prigozhin1999nonlinear,csahok2000dynamics, duran2014direct}, thought to be the only relevant length scale. However, this explanation implies that impact ripple wavelengths should tend to be larger on planets where lower density atmospheres lead to larger hoplengths, such as Mars, and smaller for thicker atmospheres with smaller hoplengths, such as Venus---in contrast with observations \cite{lapotre2016large,andreotti2021lower,miller1987wind,andreotti2006aeolian,greeley1984microdunes}. We investigate ripple emergence using a discrete element model (DEM) with a two way coupling between grains and hydrodynamics \cite{duran2012numerical}, able to reproduce observed transport characteristics and impact ripples over a wide range of atmospheric environments \cite{pahtz2020unification,pahtz2023scaling,duran2014direct, yizhaq2024coevolving} (see Supplemental Videos). Simulations show that the hoplength distribution is essentially scale-free over nearly the whole range of hoplengths $\ell$ (Fig. \ref{fig1}d; \cite{duran2014direct}). This implies that no characteristic hoplength scale relevant for ripple wavelength selection exists \cite{anderson1987theoretical}, and that our understanding of ripple formation and aeolian sediment transport is flawed. In this paper we explore the consequences of the scale-free nature of sediment transport and propose a new dynamic length scale resulting from grain-bed impacts that determines impact ripple wavelengths across planetary environments.

\begin{figure*}[t]
\centering
\includegraphics[width=\textwidth]{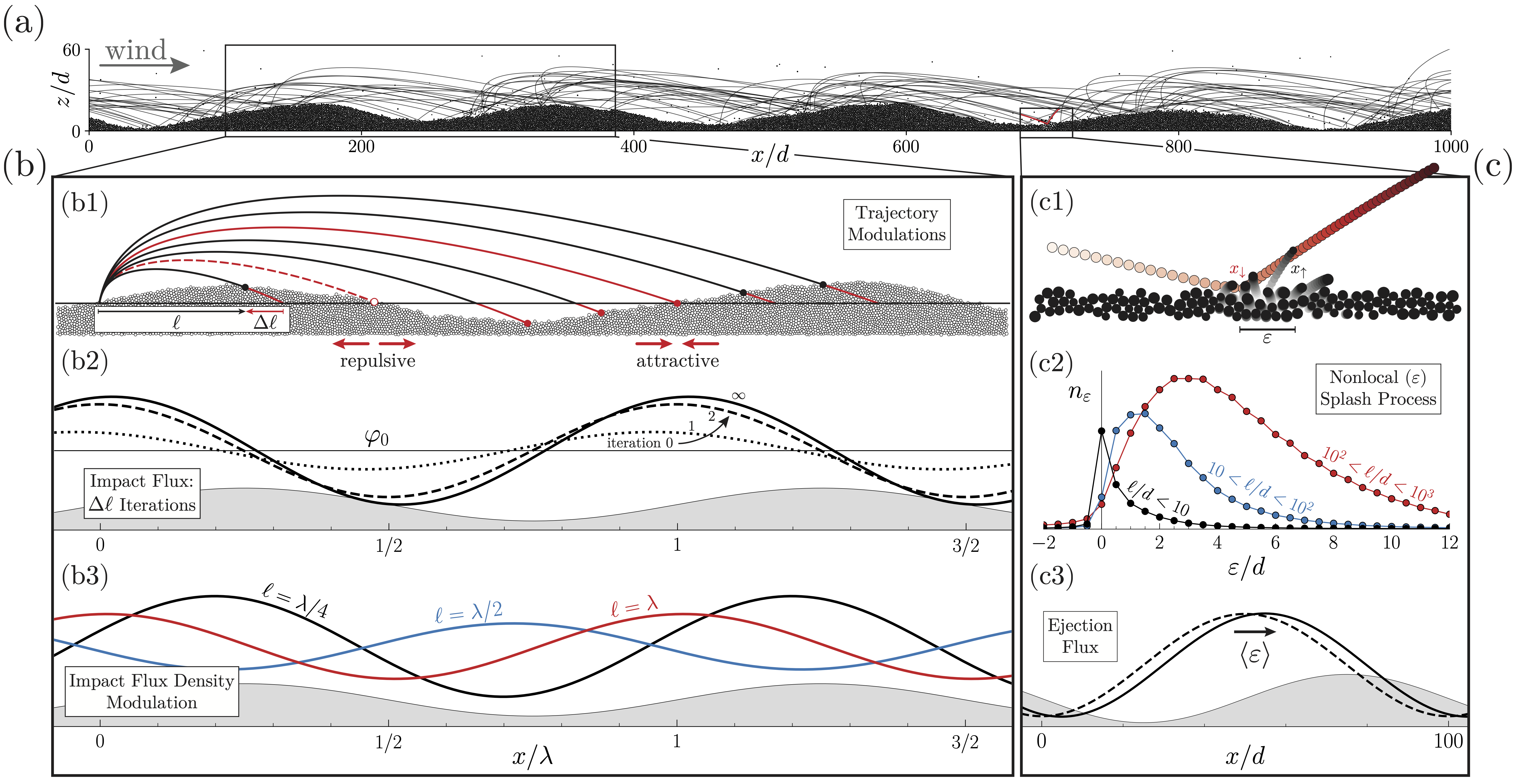}
\caption{Trajectory-ripple interactions lead to flux modulations, and impact-ejection lag phase shifts the modulated flux. (a) DEM trajectories over a rippled surface showing only hoplengths $\ell$ around one wavelength $\lambda$---illustrating the modulation of the impact flux via trajectory amplification. Panels (b) show the processes that lead to trajectory and resulting flux modulations. (b1) The mechanism for amplification and focusing of trajectories around the upwind inflection point due to geometrical modulations of hoplengths $\Delta\ell(\ell,x,t)$. Red amendments to trajectories show positive or negative changes to hoplengths $\ell$ based on where the grain impacts for a given take off location. Note that the solid red trajectory of $\ell\sim \lambda$ is the attractive in the sense that, with iterations, grains will tend towards hoplengths of $\ell\sim \lambda$. Contrary, the dashed trajectory is repulsive with iterations. Panel (b2) shows how the impact flux modulation is achieved by successive iterations of trajectories starting from a uniform impact flux $\varphi_0$ (note the increased flux at the upwind inflection point and decreased flux at the downwind inflection point). The final panel (b3) shows how the impact flux is modulated on a per hoplength basis (the impact flux density). The red line $\ell=\lambda$ reflects the impact flux density of trajectories in panel (a). The sum over all impacting hoplengths results in the total impact flux modulation as in the solid line in (b2). Panels (b) are produced for $\lambda=200d$ as in (a) which give rise to the specific phase shifts observed (see Supplementary Information). Panels (c) shows how the granular dynamics of impact lead to a granular length scale that can lead to a phase shift in the vertical flux. (c1) An impact from the DEM of a high energy grain (red) with the surface. $\varepsilon=x_\uparrow-x_\downarrow$ is the location of an ejection relative to the impact location. Panel (c2) shows DEM data for the number density of ejections and rebounds per impact of a given hoplength $n_\varepsilon(\varepsilon|\ell)$ versus rescaled $\varepsilon/d$ for various ranges of impacting hoplengths (a proxy for impact energy). The average impact lag distance $\langle\varepsilon\rangle$ has the main effect of simply phase shifting the ejection flux downwind---as illustrated in the schematic of (c3).}
\label{fig2}
\end{figure*}

\subsection*{Criticality Driven Surface Instability}

Figure \ref{fig1}d shows that grain trajectories vary continuously between the limits of very long, high-energy trajectories and more numerous very short hops, in contrast with the standard conceptual model of wind-driven transport that divides grains in motion into two distinct populations, ‘saltons’ and ‘reptons’ \cite{anderson1987theoretical,prigozhin1999nonlinear,csahok2000dynamics}. Saltons represent long trajectory, high energy grains that impact the sand surface and eject low energy reptons, which make relatively short hops. Counter to the bimodal salton-repton picture, the steady-state distribution of hoplengths, $P_0(\ell)$, measured from the DEM can be approximated as a $\sim \ell^{-1}$ power-law with an exponential cutoff for exceedingly long hoplengths set by grains that make it outside of the transport layer and whose maximum hoplengths are set by the wind speed (Fig. \ref{fig1}e). Wind tunnel observations \cite{ho2014aeolian,rasmussen2015laboratory} also report a roughly $P_0\sim \ell^{-1}$ scaling with an exponential tail (Fig. \ref{fig1}e inset). Similar predictions of $P_0(\ell)$ were also made in direct aeolian simulations and analytical modeling of \cite{lammel2017analytical}. 

The scale-free $\sim \ell^{-1}$ distribution is a characteristic signature of critical phenomena---the tendency for systems close to a threshold to be dominated by fluctuations, leading to power-law behaviors. In a classic example, on a granular slope that is gradually steepened until it approaches the angle of repose, avalanching grains can dislodge other grains poised on the edge of motion, creating cascades. Although small cascades are most numerous, these fluctuations can be arbitrarily large, exhibiting a power-law distribution of sizes \cite{held1990experimental}. In our system, the critical point may be the wind speed threshold above which grains are entrained into motion \cite{duran2011aeolian,duran2014direct}. During transport, the negative feedback of saltation on the wind strength reduces the wind speed to the threshold value at the bed \cite{duran2012numerical}. As a consequence, collisions occur on a bed of grains on the verge of motion. Thus the power-law distribution of hoplengths may result from the collisional process on an extremely sensitive grain arrangement \cite{yan2016model}. (Power-law distributions of temporal fluctuations in the number of grains in transport also suggest critical behavior \cite{held1990experimental}; Methods Fig. \ref{fig-extended data:PowerSpec}.)

Because grains impact the surface at low angles, sediment transport over ripples tends to concentrate on upwind faces, leading to an increase in ejections from the face that get distributed across the ripples crest (Fig. \ref{fig1}d, \ref{fig2}a,b). This constitutes the destabilizing mechanism that makes the ripple grow \cite{anderson1987theoretical,prigozhin1999nonlinear,csahok2000dynamics}. Additionally, the density of impacts on upwind faces tends to be iteratively amplified moving downwind (Fig. \ref{fig2}b). The amplification occurs for a subpopulation of grains leaving an upwind face with hoplengths close to one ripple wavelength (Fig. \ref{fig2}a): Hoplengths just longer than a wavelength are shortened, and shorter hoplengths are extended, depending on the relative difference in elevation between lift-off and landing (Fig. \ref{fig2}b; Methods). This focusing leads to a further concentration of impacts on the subsequent upwind face, and therefore to a further increase in the number of grains rebounding and ejected from there, completing the feedback.

 As a consequence of the scale-free hoplength distribution, the amplification of impacts on upwind slopes occurs for any wavelength perturbation, and the weighting toward small hops tends to make the smallest wavelengths grow fastest (see Supplementary Information, or SI, Section \ref{sec-supp: linear instability}). The fact that ripples in nature emerge at finite characteristic wavelengths (much greater than the grain size $d$) suggests that some mechanism stabilizes the smallest wavelengths.

\subsection*{Granular Dynamics Wavelength Selection}

Analysis of DEM results reveal a spatial lag, $\varepsilon=x_\uparrow-x_\downarrow$, between the location of an impact $x_\downarrow$ and a resulting ejection at $x_\uparrow$ (Fig. \ref{fig2}c). The momentum from an impact is distributed through a complex network of dynamic force chains \cite{owens2011sound} around the impact location, and because grains impact the bed at low angles this momentum is preferentially directed downwind. Thus, grains are often ejected several grain diameters away from an impact site---resulting in a length scale $\varepsilon$ that arises from collective granular dynamics. The spatial lag $\varepsilon$ provides the scale that determines the wavelength below which ripples cannot grow, ultimately selecting the emergent ripple size. 

The lag in the locations of ejections relative to impacts shifts the position of the erosion-deposition transition point relative to the bedform---as the ``saturation length'' does for hydrodynamic bedforms \cite{duran2019unified}. When the perturbation wavelength is sufficiently long, the lag does not play a significant role. However, when the wavelength is small enough, the lag shifts the transition point downwind of the crest, so that the crest erodes and the bedform decays. The fastest growing wavelength emerges from the competition between the destabilizing mechanism, which favors the growth of small wavelengths, and the stabilizing mechanism arising from the lag, with the result that the emergent wavelength scales with the average impact-ejection lag distance $\langle \varepsilon\rangle$ (Methods).

\begin{figure*}[t]
\centering
\includegraphics[width=\textwidth]{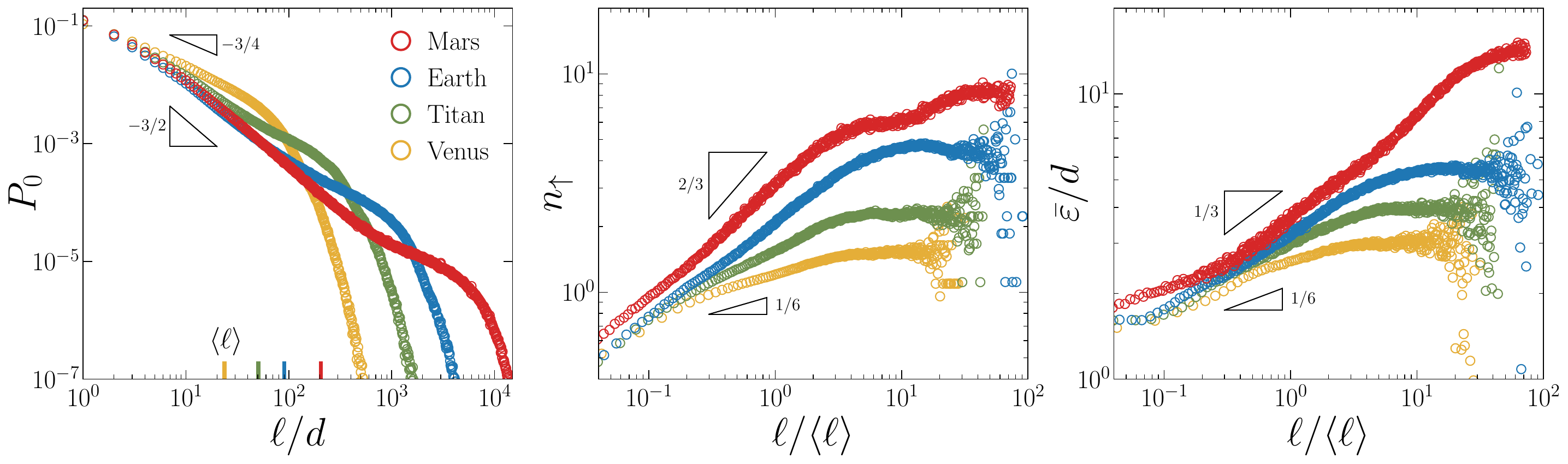}
\caption{Scale-free properties of sediment transport. The DEM suggests that the majority of trajectories ($\ell\lesssim10\langle\ell\rangle$) can be characterized by power-law functions for the hoplength distribution $P_0(\ell)$, the replacement capacity $n_\uparrow(\ell)$ (the number of ejections and rebounds per impact) and the mean impact-ejection lag distance $\bar{\varepsilon}(\ell)$---three among many other dynamical functions that show power-law behavior in the DEM. The colors show DEM data corresponding to conditions adjusted to represent Mars, Earth, Titan and Venus by varying the grain-fluid density ratio $\rho_p/\rho_f=(10^5,2000,200,50)$, respectively. All simulations are done in rarefied transport conditions with $d\simeq 100\mu$m (Methods). Inset triangles in each plot are power law references. We note that the average lag distance $\int\bar{\varepsilon}(\ell)P_0(\ell)\dd\ell\sim 2d$ for all four conditions.}
\label{fig3}
\end{figure*}

\subsection*{Analytical Model}

In the Methods and Supplementary Information (SI) we derive a generalized model for ripple growth which couples modulated grain trajectories---and resulting modulated sediment fluxes (Fig. \ref{fig2}b)---with the nonlocal impact-ejection process (Fig. \ref{fig2}c). To determine what selects initial ripple wavelengths and propagation speeds we perform a linear stability analysis to derive an expression for the ripple growth rate $\sigma(k)$ and propagation speed $c(k)$ as functions of perturbation mode $k=2\pi/\lambda$, where $\lambda$ is the perturbation wavelength (the ‘dispersion relation’ for ripple growth). This analysis reveals that ripple growth is a function of some key transport quantities measured on a flatbed: the hoplength distribution $P_0(\ell)$, the average number of ejections and rebounds per impact, or the replacement capacity $n_\uparrow(\ell)$, and the average lag distance per impact $\Bar{\varepsilon}(\ell)$, which all show scale-free/power-law behavior in the DEM for the majority of trajectories (Fig. \ref{fig3}), again hinting at the criticality of sediment transport \cite{duran2014direct,yan2016model}.

\begin{figure*}[t]
\centering
\includegraphics[width=\textwidth]{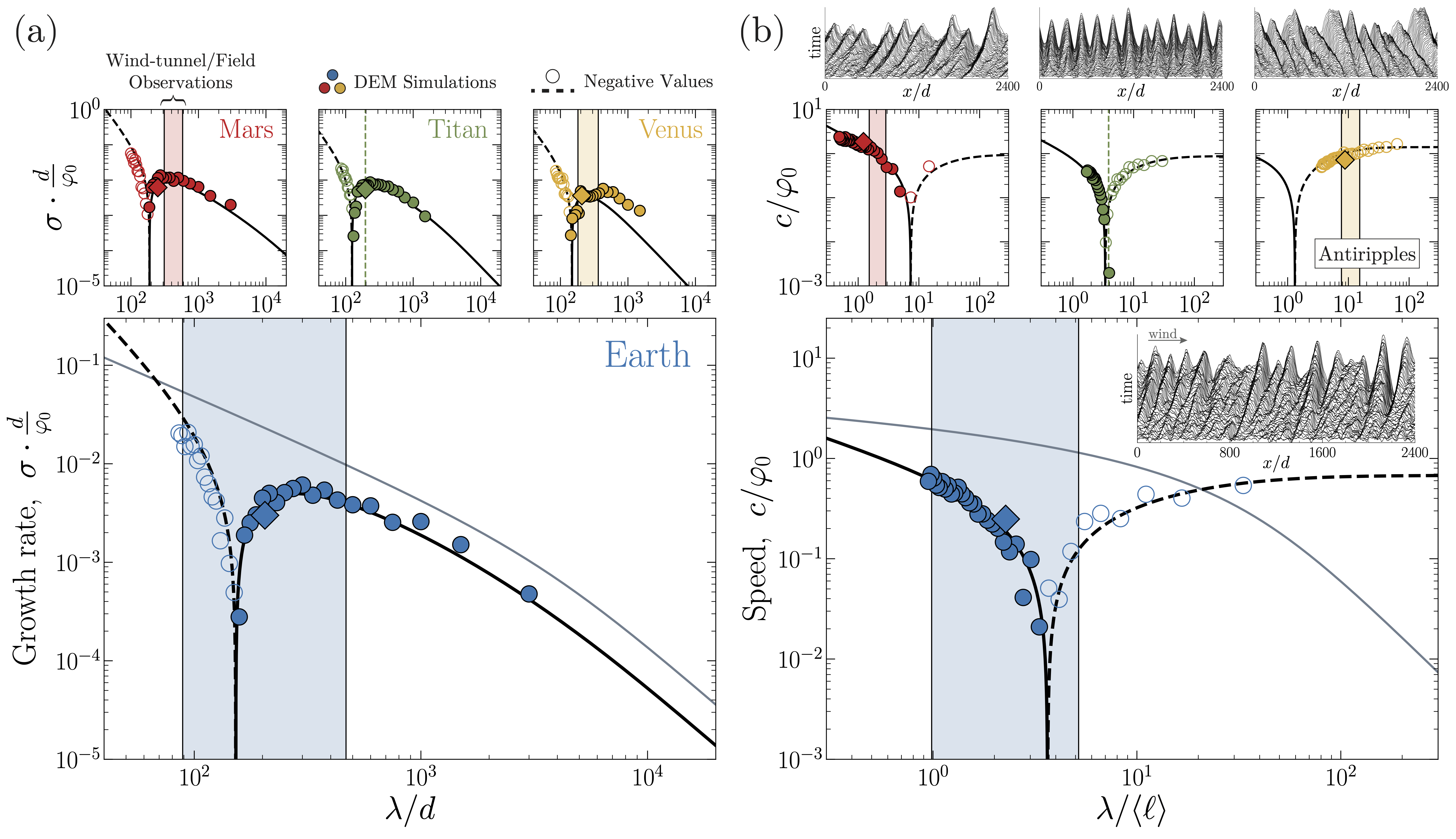}
\caption{Ripple growth rate and speed curves for Mars (red), Earth (blue), Titan (green) and Venus (gold). Circle markers are DEM experiments starting with initial bed modulations of a given wavelength $\lambda$ (see Methods). Filled circles are positive values and open circles represent negative values. Diamond markers are DEM measurements from ripple evolution from a flat initial surface condition (see space-time plots in (b)). Black lines are predictions from the new analytical model (Methods and SI) using power law approximations for input functions (Fig.'s \ref{fig3}, \ref{fig-extended data:delta ell}). Dashed black lines are negative values. (a) Ripple growth rates $\sigma$ rescaled by the homogeneous vertical flux $\varphi_0$ and the grain size $d$ versus rescaled wavelength $\lambda/d$ predict fastest growing wavelengths of roughly $\lambda\sim 200d$ for all conditions. (b) Rescaled ripple speeds $c/\varphi_0$ versus wavelength $\lambda$ rescaled by the average hoplength $\langle\ell\rangle$. Inset plots in (b) show space-time evolution of DEM ripple simulations. The grey lines for Earth in (a,b) show predictions from the salton-repton model \cite{anderson1987theoretical}, using the power-law like distribution of hoplengths $P_0(\ell)$ (SI Section \ref{sec-supp: Salton-Repton Model}). Shaded backgrounds in all plots show ranges of observed ripple sizes $\lambda$ from wind tunnel and field experiments (Mars \cite{yizhaq2024coevolving,lapotre2016large}, Earth \cite{yizhaq2024coevolving,miller1987wind,andreotti2006aeolian,walker1981experimental,schmerler2016experimental,cheng2018experimental}, Venus \cite{greeley1984microdunes}) which are in broad agreement with predictions. We note that Venus wind tunnel experiments \cite{greeley1984microdunes} did not report ripple propagation speeds---even though they were likely present---thus we can not confirm if they observed antiripples $c<0$, as predicted by our model. Additionally there is no observational data for Titan (wind tunnel or field/satellite) thus the vertical dashed line, $\lambda\sim 200d$ and $c\sim 0$, represents predictions from our model.}
\label{fig4}
\end{figure*}

In order to measure the replacement capacity $n_\uparrow(\ell)$ in the DEM \emph{during transport}, we link individual grains leaving the sand bed (crossing a horizontal plane at $z=0.7d$ from the surface) to grains that have impacted that area ($\dd x=20d$) in a time window before the crossing ($\dd t=2\sqrt{d/g}$). Values for the crossing elevation and the space and time windows are chosen to satisfy the constraint that most ejecta are paired with an impact. If the particle leaving the surface has impacted the surface within that time and space window then we count it as a rebound. Otherwise, we search for the highest energy impactor that likely gave rise to that ejecta. And $\varepsilon=x_\uparrow-x_\downarrow$ is simply the measured displacement between impact and ejection/rebound. 

Using the DEM measurements of flatbed quantities in Figure~\ref{fig3}, we find the derived dispersion relation agrees with numerical impact ripples that emerge in the DEM (Fig. \ref{fig4}, markers), and is broadly consistent with ripple observations on Earth, as well as other planets (Fig. \ref{fig4}, shaded regions). The numerical ripples produced in the DEM have been directly compared to wind-tunnel experiments for a wide range of atmospheres (from Earth to Mars-like as well as Venus-like conditions) \cite{yizhaq2024coevolving,greeley1984microdunes}. Therefore, the fact that the analytical model can reproduce the numerical ripple emergence suggests that the mechanisms accounted for in this new conceptual and analytical model capture the main physics of the system. This agreement also implies that ripple emergence is tied to the critical nature of sediment transport, because the dispersion relations of Figure \ref{fig4} (black lines) are derived using the power-law scalings observed during transport (Fig. \ref{fig3}; Methods).

In this new analytical framework, ripple wavelengths and speeds arise from the modulation of the impacting flux of grains, which strongly depends on the hoplength $\ell$ due to the focusing effect (Fig.~\ref{fig2}b). This is encoded in two dimensionless quantities $\mathcal{R}(\ell)$ and $\mathcal{S}(\ell)$ which are rescaled components of the modulated impact flux density of hoplengths in-phase and in-quadrature with respect to the topography (Methods). The in-phase modulation $\mathcal{R}(\ell)$ quantifies the contribution of trajectories of hoplength $\ell$ to the amplification of the impacting flux on ripple crests, and the out-of-phase modulation $\mathcal{S}(\ell)$ quantifies the contribution on ripple faces (e.g., Fig. \ref{fig2}b). Because these modulations result from topography-induced focusing effects (Fig. \ref{fig2}b), $\mathcal{R}(\ell)\simeq \mathcal{R}(\ell/\lambda)$ and $\mathcal{S}(\ell)\simeq \mathcal{S}(\ell/\lambda)$ primarily depend on the ratio $\ell/\lambda$ of their hoplength to the wavelength. As a first approximation $\mathcal{R} \sim \sin k\ell$ and $\mathcal{S} \sim 1 + \cos k\ell$ (Methods Fig.~\ref{fig-extended data:A}), which are maximal for $\ell \simeq\lambda/4$ (representing grains landing on the crest after being ejected on the upwind flank) and $\ell\simeq \lambda$ (representing amplified trajectories traveling from one upwind flank to the next) respectively. Our analytical model (Methods) predicts that fastest growing wavelength and associated propagation speed scale as
\begin{equation}
    \label{eq: lam and c}
    \lambda\sim \frac{\langle n_\uparrow \mathcal{S}\rangle }{\langle (1-n_\uparrow) \mathcal{R} \rangle} \langle\varepsilon\rangle \quad \text{and}\quad  c \sim \langle(n_\uparrow-1) \mathcal{S}\rangle \varphi_0,
\end{equation}
which shows that the wavelength scale is set by the ratio of the stabilizing mechanism, which scales with the average number of grains leaving from the ripple face $\langle n_\uparrow \mathcal{S}\rangle>0$ and the shift $\langle\varepsilon\rangle$, and the destabilizing mechanism, which scales with the average number of grains deposited on the crest $\langle (1-n_\uparrow) \mathcal{R} \rangle>0$ (Methods). The propagation speed scales with the homogeneous vertical flux at the surface $\varphi_0$ and the number of grains eroded from (or deposited on) the ripple faces $\langle(n_\uparrow-1) \mathcal{S}\rangle$, which may be positive \emph{or negative} depending on the ratio $\langle\ell\rangle/\lambda$ (Fig.~\ref{fig4}; Methods Fig.~\ref{fig-extended data:disp rel}), where $\langle\ell\rangle$ is the average hoplength set by the exponential cutoff of $P_0(\ell)$ (Fig. \ref{fig3}).

\subsection*{Antiripples}

Negative propagation speeds represent the novel prediction that under certain conditions impact ripples will propagate upwind as \emph{antiripples}. Upwind propagation will occur if there is net deposition on the upwind inflection point (Eq.~\ref{eq: lam and c}). When the hoplength scale is of the order of the ripple wavelength ($\langle\ell\rangle \sim \lambda$) there are many modulated high-energy trajectories (i.e., $\ell \sim \lambda$) that cause erosion of the upwind face, $\langle n_\uparrow \mathcal{S}\rangle >\langle\mathcal{S}\rangle$, and downwind propagation ($c > 0$, Eq.~\ref{eq: lam and c}). In contrast, in the limit $\langle\ell\rangle \ll \lambda$, most grains have hoplengths $\ell$ smaller than $\lambda$. Therefore on upwind slopes nearly all trajectories originate locally, and have their hoplengths shortened (Fig. \ref{fig2}b), tending to cause a reduction in the number of ejections per impact, $n_\uparrow(\ell)$ (Fig. \ref{fig3}). Thus, more grains land around the upwind inflection point than leave, $\langle\mathcal{S}\rangle>\langle n_\uparrow \mathcal{S}\rangle$, which results in net accretion on the upwind face and upwind propagation as antiripples ($c < 0$, Eq.~\ref{eq: lam and c}; Methods Fig. \ref{fig-extended data:disp rel}).

Therefore, our analysis predicts that antiripples will emerge under conditions in which the fastest growing wavelength, which is ultimately selected by the granular length scale $\langle\varepsilon\rangle$, is much larger than the hoplength scale $\langle\ell\rangle$. This is predicted to occur when the density ratio between the grains and the fluid $\rho_p/\rho_f$ is small and transport is tending to transition to bedload (smaller hoplengths), as on some other planetary bodies such as Titan and Venus (Fig. \ref{fig4}b), and may also occur on Earth for gravel sized grains where the hoplengths are also sufficiently short.

\subsection*{Implications}

Impact ripples occur across the globe and solar system. In our new model for the emergence of these patterns, we find that rather than ripple wavelengths being selected by a sediment transport length scale, they are selected by a bed deformation length scale, $\langle\varepsilon\rangle$. This length scale is analogous to the impact-ejection lag length in nanometer-scale ion sputtering (erosion) processes which selects the wavelengths of micron-scale ripples on surfaces bombarded by ions (at low temperatures) \cite{makeev2002morphology,norris2019ion}. Additionally, the mechanism that allows $\langle\varepsilon\rangle$ to determine wavelengths that are orders of magnitude larger (via scale-free impact forcing) is analogous to the amplification of the sediment-flux saturation length $L_{\mathrm{sat}}$ which selects dune sizes (scale-free fluid forcing \cite{charru2013sand,duran2019unified}).

The importance of the length scale $\varepsilon$ provides insights into long standing problems of ripple sizes on other planetary bodies. Because this bed deformation length is a granular phenomenon dependent mostly on grain properties, such as $d$, and depending little on the on the structure of the fluid or atmospheric density, we may expect that ripple wavelengths on other planets with the same grain composition should have roughly the same impact ripple sizes---which is what we find when comparing Earth to Venus and Martian impact ripples (Fig. \ref{fig4}; \cite{lapotre2016large,yizhaq2024coevolving,greeley1984microdunes}). 

While wavelengths are selected by grain-bed interactions, ripple propagation speeds are almost uniquely determined by the vertical flux scale $\varphi_0$---set by the wind speed---and the scale of the hoplength distribution $\langle\ell\rangle$---set by the atmospheric density and grain density and size. This suggests that while wavelengths may not vary much from planet to planet, propagation speeds may vary significantly---even from downwind to upwind propagating \emph{antiripples} (Fig. \ref{fig4}b). Predictions of negative speed antiripples on Earth, with sufficiently coarse or light grains, can be tested in wind tunnels and field experiments. 

Subsequent work should address how wind modulation affects the formation of impact ripples, as well as the convolution of impact and hydrodynamically forced bedforms \cite{yizhaq2024coevolving}. Additionally, pattern coarsening from nonlinear 3-dimensional dynamics over long time scales modify the ripples that initially emerge from the 2-dimensional linear dynamics we focus on here \cite{werner1993fundamentally,landry1994computer, huntley2008influence}. Although the density of defects in the initial pattern determines how much wavelength can increase \cite{werner1999bedform,huntley2008influence}, wavelengths typically only coarsen by approximately a factor of two or so based on wind tunnel observations \cite{andreotti2006aeolian,andreotti2021lower} and the distribution of ripple wavelengths observed on Earth and Mars \cite{lapotre2016large}. Thus it is unlikely that coarsening effects are significant enough to erase the memory of the linear instability. Additionally, previous modeling \cite{duran2014direct} and experimental \cite{andreotti2006aeolian} work indicates that the initially fastest growing wavelength can increase moderately with wind speed.  Wavelengths increase when wind speed,  sediment flux, and impact densities are high enough that a continuously excited `gaseous layer' forms at the bed surface \cite{duran2014direct}. Impact ripples can form for wind speeds only slightly above the threshold for sediment transport, when impacts are rarefied enough that no gaseous layer exists.  We have focused here on conditions near the threshold, to explore the most general, fundamental mechanisms causing ripple formation. Possible future work could derive parametrizations analogous to those in Fig. 2 for higher wind speed, enabling an investigation of the relationship between wavelength and wind speed  using our analytical model (Methods and SI).

Future efforts to compare model predictions to observations will test the new model of ripple formation, but more importantly they will also implicitly test the implications that scale-free transport dynamics (Fig. \ref{fig3}) and granular length scales are important fundamental aspects of aeolian sediment transport, generally.

\newpage

\setcounter{equation}{0}
\setcounter{figure}{0}
\setcounter{table}{0}
\makeatletter
\renewcommand{\theequation}{M\arabic{equation}}
\renewcommand{\thefigure}{M\arabic{figure}}
\renewcommand{\thetable}{M\arabic{table}}

\begin{center}
\textbf{\large Methods}
\end{center}

\subsection*{Surface Evolution}\label{sec-methods: surface evo}
To quantitatively explore the conceptual model described in the main text, we consider the evolution of the surface elevation $Z(x,t)$ using the conservation of mass, which can be alternately expressed in terms of gradients in horizontal flux or in terms of vertical fluxes (SI \ref{sec-supp: Surface Evolution}): 
\begin{equation}
    \label{eq-methods: dZ/dt = -dQ/dx}
    \partial_t Z(x,t) = \varphi_\downarrow(x,t)-\varphi_\uparrow(x,t),
\end{equation}
where $\varphi_\downarrow(x,t)$ and $\varphi_\uparrow(x,t)$ are the vertical fluxes of grains respectively reaching the bed and leaving the bed, normalized by the bed packing fraction $\phi_b$. 

\subsection*{Linking Impacts to Ejections}\label{sec-methods: Linking Impacts to Ejections}
 The impact flux density $\psi(\ell, x_\downarrow)$ is defined such that $\psi(\ell, x_\downarrow)\dd\ell \dd x_\downarrow$ is the volume of the grains (packed at the bed volume fraction $\phi_b$) arriving per unit time in the interval $[x_\downarrow, x_\downarrow + \dd x_\downarrow]$ after a hop of length between $\ell$ and $\ell + \dd\ell$. It can be expressed as a function of the distribution of the hoplengths  $\ell$ on a flatbed, noted $P_0 (\ell)$ (Fig. \ref{fig1}e, \ref{fig3}), and of the hoplength modulation, noted $\Delta\ell(\ell,x_\downarrow)$, which result from surface perturbations (Fig. \ref{fig2}b). From the conservation of mass, $\psi(\ell,x_\downarrow)$ can be linked to upwind ejections/rebounds, $\varphi_\uparrow(x_\downarrow-\ell)$,  by the expansion (SI \ref{sec-supp: link impacts to ejections})
\begin{eqnarray}
    \label{eq-methods: psi(l,x)}
    \psi(\ell,x_\downarrow)\sim\varphi_\uparrow(x_\downarrow-\ell)P_0(\ell) -\varphi_0 (\partial_{\ell}+\partial_{x_\downarrow})\Delta\ell(\ell,x_\downarrow) P_0(\ell)
\end{eqnarray}
where  $\varphi_0$ is the vertical flux in the homogeneous and steady-state, for which $\varphi_\downarrow(x)=\varphi_\uparrow(x)= \varphi_0$ and $\psi(\ell,x)=\psi_0(\ell) = \varphi_0P_0(\ell)$. Stated simply, the impact flux density of grains is the translation of the ejection flux of grains $\varphi_\uparrow(x_\downarrow-\ell)$ traveling a distance $\ell=x_\downarrow-x_\uparrow$ determined by the homogeneous distribution $P_0(\ell)$, plus the flux modulation associated with the changes in trajectories $\Delta\ell(\ell,x_\downarrow)$. Integrating this equation over $\ell$ links the total impact flux $\varphi_\downarrow$ to the vertical flux $\varphi_\uparrow$ by adding up the flux contributions from all trajectories:
\begin{eqnarray}
    \label{eq-methods: varphi_down(x)}
    \varphi_\downarrow(x_\downarrow)&=&\int \psi(\ell,x_\downarrow)\dd\ell\\
 &=&   \int \varphi_\uparrow(x_\downarrow-\ell)P_0(\ell)\dd\ell - \varphi_0\partial_{x_\downarrow}\langle \Delta\ell(x_\downarrow)\rangle\nonumber
\end{eqnarray}
The second term involves the spatial gradient of the average change in hoplengths $\langle \Delta\ell(x_\downarrow)\rangle$  (SI \ref{sec-supp: linear regime}). 

\subsection*{Hoplength Modulations}\label{subsec-methods: Hoplength Modulations}

In the presence of a modulated surface, the hoplength $\ell$ is not in general the same as the flat bed hoplength $\ell_0$ and can be written as 
\begin{equation}
\ell = \ell_0 + \Delta\ell, 
\end{equation}
where hoplength changes $\Delta\ell(\ell,x,t)$ are a function of trajectory interactions with surface geometry (at lift-off $x_\uparrow$ and on impact $x_\downarrow$) and wind modulation effects over the surface, in general. On impact, the geometrical changes in hoplengths from surface perturbations are given by (with reference to Fig. \ref{fig2}b)
\begin{equation}\label{eq-methods: Dl}
\Delta\ell(\ell,x_\downarrow)=\cot\theta(\ell)[Z(x_\uparrow)-Z(x_\downarrow)],
\end{equation}
for impact angle $\theta(\ell)$. The behavior of geometrical $\Delta\ell$ over a modulated surface is shown in Figure \ref{fig-extended data:delta ell}.  These geometrical effects on $\Delta\ell$ tend to cause impact focusing around the upwind inflection ripple point and is the main contributor to the flux modulations and $\ell\sim\lambda$ amplification---as illustrated in Figure \ref{fig2}b in the main text. The concept of hoplength changes $\Delta\ell$ has been used in past impact ripple models \cite{nishimori1993formation,prigozhin1999nonlinear}, most notably in the limit of very small hoplengths where $\Delta\ell\sim -\cot\theta\ell\partial_xZ$ \cite{prigozhin1999nonlinear}, but never in its general form as derived here. We also note that $\cot\theta(\ell)$ is another function measured in the DEM that shows approximately power-law like behaviour for the majority of particles in transport $\ell\lesssim \langle\ell\rangle$ (Fig. \ref{fig-extended data:delta ell}).

\subsection*{Linking Ejections to Impacts}\label{subsec-methods: Linking Ejections to Impacts}
Given that Equation \eqref{eq-methods: varphi_down(x)} expresses the impact flux $\varphi_\downarrow$ in terms of the ejection flux $\varphi_\uparrow$ and hoplength modulations $\Delta\ell$ (Eq. \ref{eq-methods: Dl}), to close the problem we must relate ejections to impacts. We do this using a novel \emph{nonlocal} splash-process approach. For an ensemble of impacts of a given hoplength $\ell$ at impact site $x_\downarrow$ there are on average $n_\varepsilon(\varepsilon|\ell)$ grains that are ejected at distance $\varepsilon=x_\uparrow-x_\downarrow$ away from the impact site, per unit distance (as seen in the DEM (Fig. \ref{fig2}c) and experiments \cite{ammi2009three}). The rate of ejections and rebounds is given by
\begin{eqnarray}
    \label{eq-methods: varphi_uparrow(x)}
    \varphi_\uparrow(x_\uparrow) &=&\iint n_\varepsilon(\varepsilon|\ell)\psi(\ell,x_\uparrow-\varepsilon)\dd \varepsilon \dd \ell \\ 
    &\simeq& \int n_\uparrow(\ell)\psi(\ell,x_\uparrow) \dd \ell - \partial_{x_\uparrow} \int \bar{\varepsilon}(\ell)n_\uparrow(\ell)\psi(\ell,x_\uparrow) \dd \ell \notag,
\end{eqnarray}
where the total number of ejections and rebounds per impact (the replacement capacity) is $n_\uparrow(\ell)=\int n_\varepsilon(\varepsilon|\ell)\dd\varepsilon$ and the average impact-ejection lag distance per impact is $\bar{\varepsilon}(\ell)=n_\uparrow(\ell)^{-1}\int \varepsilon n_\varepsilon(\varepsilon|\ell)\dd\varepsilon$. In general $n_\uparrow(\ell)$ and $\bar{\varepsilon}(\ell)$ are functions of impact energy and other quantities, however, in the DEM we find the hoplength to be a good proxy for impact energy. 

Equation \eqref{eq-methods: varphi_uparrow(x)}, evaluated in the homogeneous steady state, implies that the replacement capacity $n_\uparrow(\ell)$ is constrained to be 
\begin{equation}\label{eq-methods: <n>=1}
    \langle n_{\uparrow}\rangle \equiv 1.
\end{equation}
This useful constraint indicates that $n_{\uparrow}(\ell)P_0(\ell)=P_n(\ell)$ can be thought of as a probability distribution (SI \ref{sec-supp: linear instability}). 

The constraint that $\langle n_{\uparrow}\rangle=\int n_\uparrow (\ell)P_0(\ell)\dd\ell=1$ for \emph{all} steady-state transport conditions also implies that $\langle \varepsilon\rangle=\int \bar\varepsilon(\ell) P_0(\ell)\dd\ell$ varies little across transport conditions. Figure \ref{fig3} suggests that $n_\uparrow\sim \Bar{\varepsilon}^2$ for large density ratios $\rho_p/\rho_f$ (Mars, Earth) suggesting that $\Bar{\varepsilon}^2$ is the (2D) volume of grains ejected from the surface. And for small $\rho_p/\rho_f$ (Venus), $n_\uparrow\sim \Bar{\varepsilon}$ representing a more shear-like dislodging of surface grains. This behavior is characteristic of aeolian saltation and bedload transport, respectively \cite{pahtz2017fluid}. Such relationships between $n_\uparrow(\ell)$ and $\bar\varepsilon(\ell)$ suggests that can that the average $\langle\varepsilon\rangle\propto d$ is roughly constant across planetary conditions---which is what we find in the DEM (see caption of Fig. \ref{fig3}).

\subsection*{Linear Instability: Wavelength and Speed Selection}\label{subsec-methods: Linear Instability: Wavelength and Speed Selection}
Equations (\ref{eq-methods: dZ/dt = -dQ/dx}-\ref{eq-methods: varphi_uparrow(x)}) provide a closed form model for surface evolution, given the flatbed quantities $P_0(\ell),\ n_\uparrow(\ell),\ \bar{\varepsilon}(\ell)$ and $\cot\theta(\ell)$. To understand the dynamics of ripple emergence from a flat surface, we perform a linear stability analysis of the flat bed solution. Small perturbations to the surface $Z(x,t)$, such that $|\partial_xZ|\ll1$, can be decomposed in Fourier space, $Z_k(x,t)=\hat{Z}(t)e^{ikx}$ (SI \ref{sec-supp: Growth and Modulation Rates}). The mode $k=2\pi/\lambda$ grows exponentially in time according to 
\begin{equation}
\hat{Z}(t) \sim \hat{Z}(0)e^{(\sigma -ikc)t}
\end{equation}
where $\sigma(k)$ is the growth rate and $c(k)$ the propagation speed.

The Fourier transform of the impact flux density obeys at the linear order 
\begin{equation}
    \hat{\psi}(\ell,t)\sim \psi_0(\ell) (\mathcal{R}+i\mathcal{S})\;k\hat{Z}(t)
\end{equation}
for $\mathcal{R}(\ell)$ and $\mathcal{S}(\ell)$ are the in-phase and out-of-phase components the impact flux amplifications respectively, relative to the modulated surface, and are approximately unique functions of $\ell/\lambda$---i.e., scale invariant (Fig.'s \ref{fig2}b and \ref{fig-extended data:A}). The growth rate and speed of a given perturbation mode deduce as:
\begin{eqnarray}
\label{eq-methods: Omega(k) = <R> -<S>}
    \frac{\sigma}{k\varphi_0} &\sim& \langle(1-n_\uparrow)\mathcal{R}\rangle - k\langle\varepsilon_{\Delta\ell}\rangle_n\langle n_\uparrow\mathcal{S}\rangle  \notag\\
     \frac{c}{\varphi_0} &\sim &\langle(n_\uparrow-1)\mathcal{S}\rangle
\end{eqnarray}
where $\langle\varepsilon_{\Delta\ell}\rangle_n$ is a weighted average of $\bar{\varepsilon}(\ell)$ over hoplength changes $\Delta\ell(\ell)$---i.e., for geometrical hoplength changes (Eq. \ref{eq-methods: Dl}) $\langle\varepsilon_{\Delta\ell}\rangle_n=\langle\varepsilon_{\theta}\rangle_n=\langle\bar\varepsilon\cot\theta\rangle_n/\langle\cot\theta\rangle_n$ where $\langle\cdot\rangle_n=\int(\cdot )P_n\dd\ell$  (see SI Section \ref{sec-supp: linear instability}). This way of stating the dispersion relation has a particularly intuitive meaning: The destabilizing mechanism represents the average number of grains that remain on the crest after traveling from the upwind face $\langle(1-n_\uparrow)\mathcal{R}\rangle$. This effect is countered by the stabilization mechanism from the increase number of grains being eroded from the crest due to the down wind shift in the ejection flux from $k\langle\varepsilon_{\Delta\ell}\rangle_n\langle n_\uparrow\mathcal{S}\rangle$. And the propagation speeds are determined by the number of grains eroded from or deposited on the upwind face $\langle(n_\uparrow-1)\mathcal{S}\rangle$ which is positive in the regime where the wavelength $\lambda$ is of order $\langle \ell \rangle$ because grains are being distributed from the upwind face over to the back of the ripple, and negative for wavelengths in the regime $\lambda\gg\langle\ell\rangle$.

The fastest growing wavelength of Equation \eqref{eq-methods: Omega(k) = <R> -<S>} is found at the maximum $\partial_k\sigma=0$ which gives the scaling approximation provided in Equation \eqref{eq: lam and c} in the main text. Note that in Equation \eqref{eq: lam and c} in the main text we use $\langle\varepsilon_{\Delta\ell}\rangle_n\equiv \langle\varepsilon\rangle$ for simplicity. The wavelength and speed scalings from this model (Eq.'s \ref{eq-methods: Omega(k) = <R> -<S>} and \ref{eq: lam and c}) can be seen in Figure \ref{fig-extended data:disp rel} over a wide range of transport conditions.

\subsection*{DEM Simulations}\label{subsec-methods: DEM Simulations}

The discrete element model (DEM) used in this study to simulate sediment transport and ripple formation resolves grain-fluid and grain-grain forces on each individual particle. The simulations are quasi-3D with an along wind direction, $x$, of length $3000d$ with periodic boundaries, a vertical direction, $z$, of height $800d$ (higher than the largest hopheight), and a cross wind direction $y=1d$. The domain is filled with $\sim 30,000$ grains of diameter $d$ with minor polydispersity. See Duran et al., 2012 \cite{duran2012numerical}, for a detailed description of the model dynamics.

In general, sediment transport is driven by fluid drag on sand grains. Thus the system is chiefly controlled by six parameters: the particle density $\rho_p$ and diameter $d$, the fluid density $\rho_f$, kinematic viscosity $\nu$, and shear velocity $u_*$, and gravity $g$. With these sediment transport can be characterized by three dimensionless numbers,
\begin{equation}
    \frac{\rho_p}{\rho_f},\quad \text{Ga}=\frac{\sqrt{(\rho_p/\rho_f-1)gd^3}}{\nu},\quad \Theta=\frac{u_*^2}{(\rho_p/\rho_f-1)gd},
\end{equation}
which are used to vary the transport conditions in the DEM. The density ratio between particles and the fluid $\rho_p/\rho_f$ is used to set the effective planetary regime, the Galileo number Ga is used to set the grain size $d$, and the Shields number $\Theta$ sets the wind speed. Table \ref{table-methods: DEM params} shows the values of these three numbers used in the DEM simulations presented in this study.

In our planetary simulations, ripples emerge spontaneously from flatbed, steady-state transport as seen in the space-time profiles in Figure \ref{fig4}b and in the Supplemental Video. Additionally, for each flatbed simulation we also conducted a range of simulations with the initial bed modulated as a sine wave of a given wavelength $\lambda$ and small amplitude ($\lesssim 3d$, much smaller than the imposed wavelength $\lambda$) to directly simulate the linear instability and measure the growth rates $\sigma(k)$ and speeds $c(k)$ for various perturbation modes $k=2\pi/\lambda$ (as was done in \cite{duran2014direct}). As seen in Figure \ref{fig4}, the fastest growing wavelength measured by direct simulations of $\sigma(k)$ agrees with the wavelength that spontaneously emerges from the flatbed simulations (diamond markers).

In order to obtain a closed analytical solution for $\sigma(k)$ and $c(k)$ based on our model derived above we approximate the flatbed distribution of hoplengths $P_0(\ell)$ as a power law with an exponential tail
\begin{equation}\label{eq-methods: P0}
    P_0\sim \ell^{-1+\alpha}e^{-\ell/\mu},
\end{equation}
where $\alpha$ takes into account the small deviation from the $\sim\ell^{-1}$ scaling observed for different density ratios (Fig. \ref{fig3}) and $\mu\sim \langle\ell\rangle$ sets the scale of the distribution---values of $\mu$ are chosen such that the analytical values of $\langle\ell\rangle$ matches the DEM measurements. For the normalization of $P_0$, $\int_{\ell_i}^{\infty}\ell^{-1+\alpha}e^{-\ell/\mu}\dd\ell$ we use a minimum hoplength of $\ell_i=0.1d$ for $\alpha\leq 0$ and $\ell_i=0$ for $\alpha>0$. We also approximate the other remaining functions used in the analytical model as power laws of the form
\begin{equation}\label{eq-methods: nup, eps, cot0}
    n_\uparrow\sim \ell^{\gamma_n}/\langle \ell^{\gamma_n}\rangle,\quad \bar\varepsilon\sim \bar\varepsilon_{1d}(\ell/d)^{\gamma_\varepsilon},\quad \cot\theta\sim \cot\theta_{1d}(\ell/d)^{\gamma_\theta}
\end{equation}
where $\langle \ell^{\gamma_n}\rangle$ comes from the constraint $\langle n_\uparrow\rangle=1.$ The values $\bar\varepsilon_{1d}$ and $\cot\theta_{1d}$ are the quantities when the hoplength $\ell=1d.$ Parameters used for calculation of the dispersion relation predictions in Figure \ref{fig4} are seen in Table \ref{table-methods: DEM params}. 

Direct measurements of $\bar\varepsilon(\ell)$ in the DEM during transport is complicated by the fact that we must measure $\bar\varepsilon$ for grains crossing a certain level ($0.7d$) above the average surface elevation and the fact the the surface geometry is not smooth. The functional form of $\bar\varepsilon$ measurements seem to be robust when compared to single particle collision experiments (not shown) but measurement error is observed during transport, mostly for small hoplengths. Thus we fix $\bar\varepsilon_{1d}=1d$ for all dispersion relation predictions---a physically reasonable constraint---and just fit the power $\gamma_\varepsilon$ (Table \ref{table-methods: DEM params}). The other parameters used in the model are approximated from the DEM measurements seen in Figures \ref{fig3} and \ref{fig-extended data:delta ell}. 

We lastly note that the model predictions of the dispersion relation ($\sigma$ and $c$) have been scaled by a proportionality factor to provide best fits for DEM measurements (Fig. \ref{fig4}). We do not expect the DEM measured dispersion relation to match the predictions exactly for three reasons: 1) the vertical flux scale $\varphi_0$ is measured in the DEM at $0.7d$ above the surface and should therefore be smaller than the theoretical value of $\varphi_0$ at the surface, 2) the analytical  fitting functions of DEM quantities in Equations \eqref{eq-methods: P0} and \eqref{eq-methods: nup, eps, cot0} are approximations, and 3) the hoplength changes $\Delta\ell$, which set the flux modulation magnitude (and therefore growth rate and speed magnitudes; SI \ref{sec-supp: Growth and Modulation Rates}), used in our analytical predictions consist of only geometrical contributions (Eq. \ref{eq-methods: Dl}) and are therefore smaller in magnitude than what is measured in the DEM. Namely, the DEM also simulates $\Delta\ell$ contributions from ejection angles responses to surface geometry---which would increase the overall magnitude of $\Delta\ell$ and add increasing flux modulation effects for small $\lambda$ (higher slopes). These additional effects are not the main contributors of the flux modulations (Fig. \ref{fig-extended data:A}) and are thus effectively ignored, up to a constant, in this version of the analytical model. For these three reasons we introduce a multiplicative constant to fit the predicted dispersion relations for planetary condition simulated in the DEM. This constant is labeled as $\sigma_\text{DEM}/\sigma_\text{pred}$ in Table \ref{table-methods: DEM params}. Future work will be done to better understand the consequences more precise analytical approximations (Eq.'s \ref{eq-methods: P0} and \ref{eq-methods: nup, eps, cot0}) and of additional contributions to hoplength modulations $\Delta\ell$ from ejection angle and wind modulation effects.

\textbf{Author Contributions:} All authors contributed to experimental design, interpretation and writing.\vspace{0.5cm}

\textbf{Data Availability Statement:} Data will be available in an external repository after paper acceptance. \vspace{0.5cm}

\textbf{Code Availability Statement:} Code for the discrete element model (DEM) used in this study can be made available upon request. \vspace{0.5cm}

\textbf{Competing Interest Statement:} The authors declare no competing interests.


\newcolumntype{C}{>{\centering\arraybackslash}X|}
\renewcommand{\arraystretch}{2}
\begin{table}[t]
\centering
    \begin{center}
        \caption{DEM simulation parameters and associated fit values for the analytical model. The values of Ga are chosen to keep $d\simeq 100\mu$m and the values of $\Theta$ are chosen such that the steady-state number of particles in transport per unit area is $\sim 1/50d^2$---representing fairly rarefied transport conditions.}\label{table-methods: DEM params}
        \begin{tabularx}{\textwidth}{|CCCC|CCCCCCCCC}
        \hline
        \quad & $\rho_p/\rho_f$ & Ga & $\Theta$ & $\mu/d$ & $\alpha$ & $\gamma_n$ & $\gamma_\varepsilon$ & $\gamma_\theta$ & $\bar\varepsilon_{1d}/d$ & $\cot\theta_{1d}$ & $\frac{\sigma_{\text{DEM}}}{\sigma_{\text{pred}}}$\\ \hline\hline
        Mars & $10^5$ & 2 & 0.017 & $4800$ & $-0.15$ & $0.52$ & $0.27$ & $0.24$ & $1$ & $0.95$ & $0.29$ \\ \hline 
        Earth & $2000$ & 10 & 0.017 & $750$ & $0$ & $0.4$ & $0.25$ & $0.24$ & $1$ & $0.95$ & $0.51$ \\ \hline
        Titan & $200$ & 14 & 0.032 & $393$ & $0$ & $0.35$ & $0.23$ & $0.24$ & $1$ & $1.5$ & $0.54$ \\ \hline
        Venus & $50$ & 40 & 0.036 & $63$ & $0.38$ & $0.18$ & $0.15$ & $0.07$ & $1$ & $3$ & $0.29$ \\ \hline
        \end{tabularx}
    \end{center}
\end{table}

\begin{figure*}[b]
    \centering
    \includegraphics[width=\textwidth]{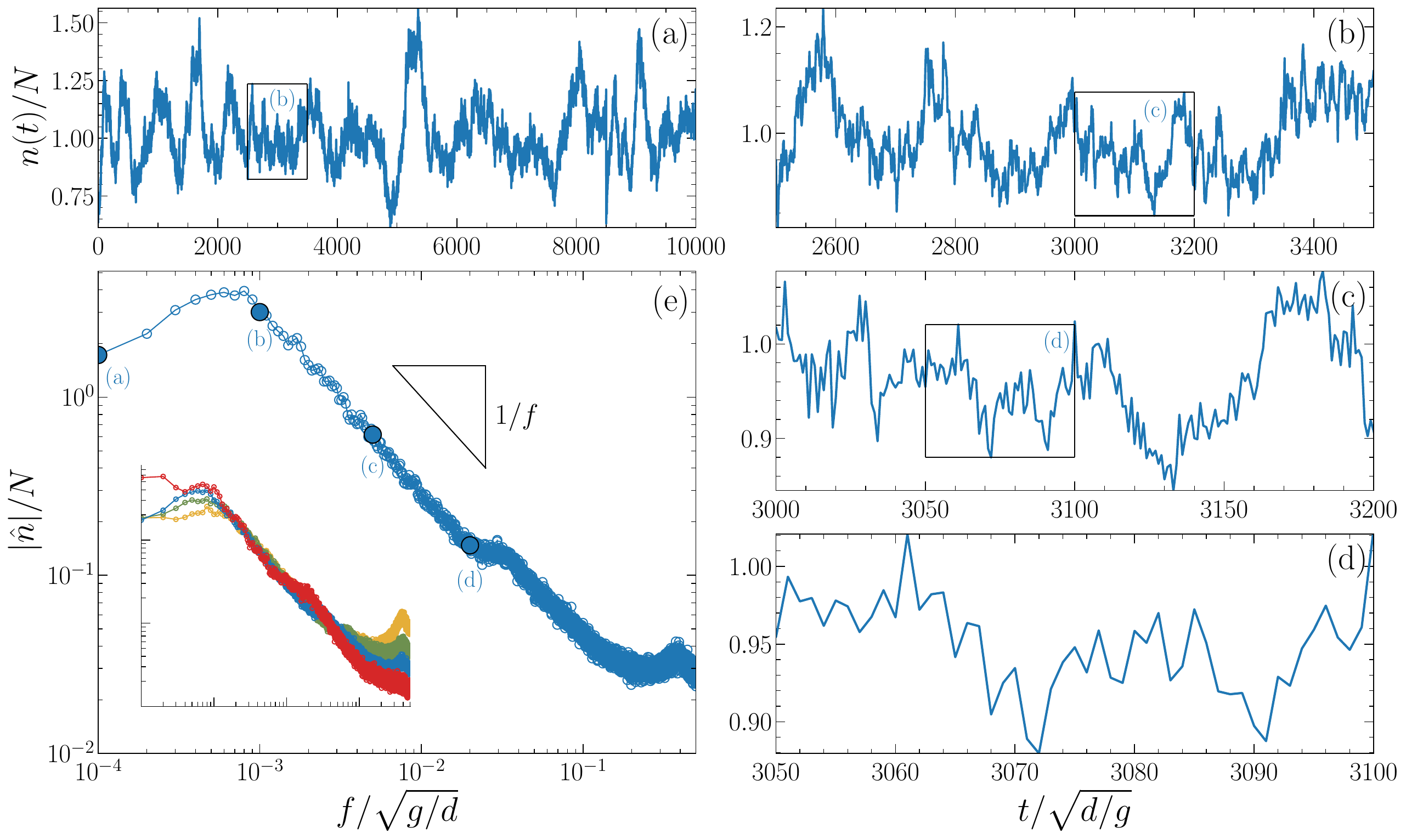}
    \caption{(a-d) DEM simulations of transport on Earth showing the time series of the number of grains in transport per unit area $n(t)$ rescaled by the average $N\sim 1/50d^2$. Panel (b) is the zoomed in box outlined in (a) and (c) is the black box in (b) and similarly (d) is the zoomed box of (c). Panel (e) shows the spectrum of fluctuations in $n(t)$ of a given frequency $|\hat{n}(f)|$ (i.e., the square root of the power spectrum). The spectrum shows power law decay as $1/f$ hinting at the critical sensitivity of the granular surface: With the wind speed at the bed buffered to the threshold of motion \cite{duran2011aeolian,duran2014direct}, the bed is sensitive to perturbations. In this condition, when an impact ejects grains that then eject other grains, and so on, this feedback can cascade into large bursts of transport. These cascades cause fluctuations in transport across a range of scales. Highlighted by the solid markers in (e) are the frequency windows used in (a-d). Note that (b-d) fall in the scale-free region. The insets in (e) show the spectrum for Mars (red), Earth (blue), Titan (green) and Venus (gold) all showing similar scale-free $|\hat{n}|\sim 1/f$ spectra.}
    \label{fig-extended data:PowerSpec}
\end{figure*}

\begin{figure*}[b]
    \centering
    \includegraphics[width=\textwidth]{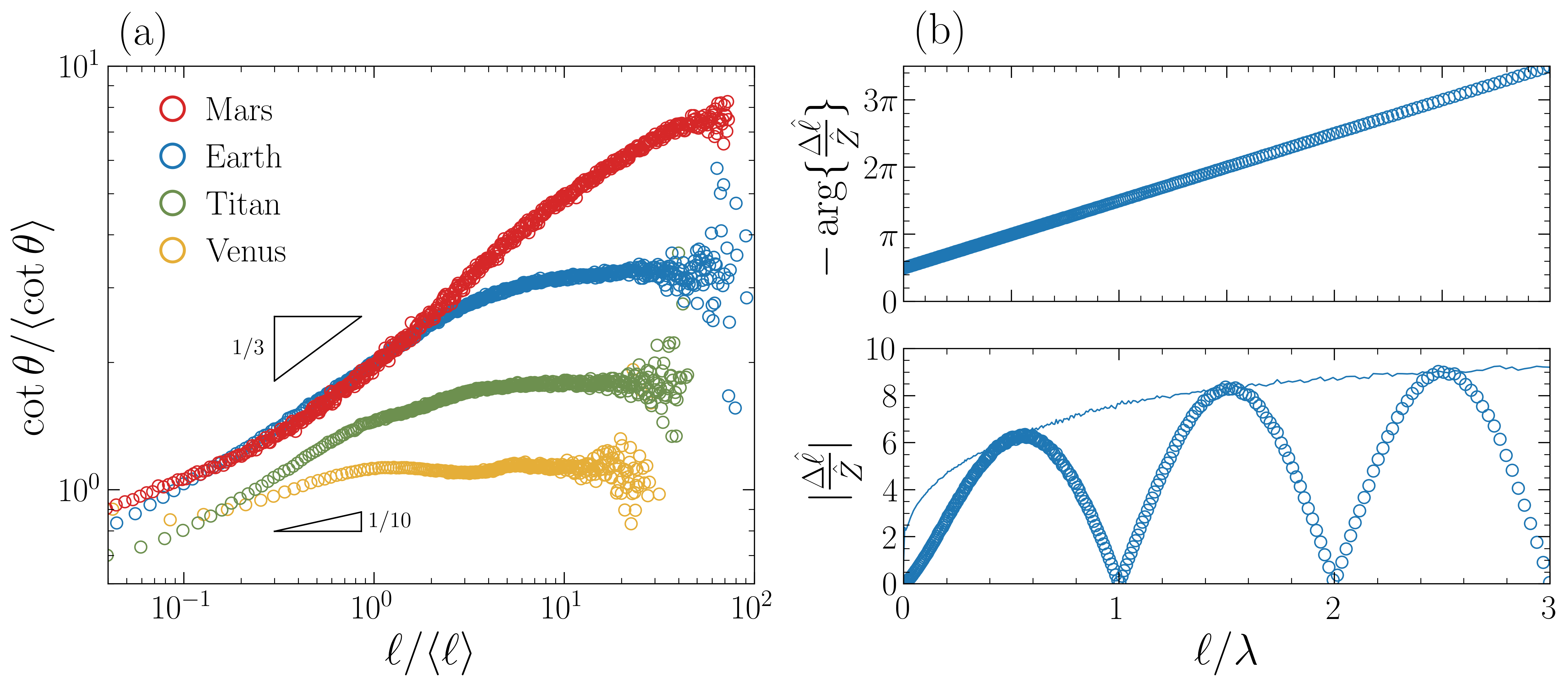}
    \caption{(a) Cotangent of impact angles $\cot\theta(\ell)$ for Mars (red), Earth (blue), Titan (green) and Venus (gold) conditions, measured from DEM simulations. $\cot\theta(\ell)$ is rescaled by its average for clarity of presentation. For Mars, Earth, Titan and Venus $\langle\cot\theta\rangle\simeq(1.4,\ 1.5,\ 2.8,\ 3.7)$, respectively. (b) The Fourier transform of geometric hoplength modulations (for small surface perturbations $\hat{Z}$) are defined as $\Delta\hat{\ell}=\cot\theta(\ell)(e^{-ik\ell}-1)\hat{Z}$ (Eq. \ref{eq-methods: Dl}). Plotted in (b) is the (negative) phase and magnitude of $\Delta\hat{\ell}/\hat{Z}$ versus rescaled hoplength $\ell/\lambda$ for $\lambda=200d$ in Earth conditions ($\langle\ell\rangle\simeq 100d$). Also shown over the magnitude is the envelop $2\cot\theta(\ell)$. }
    \label{fig-extended data:delta ell}
\end{figure*}

\begin{figure*}[b]
    \centering
    \includegraphics[width=\textwidth]{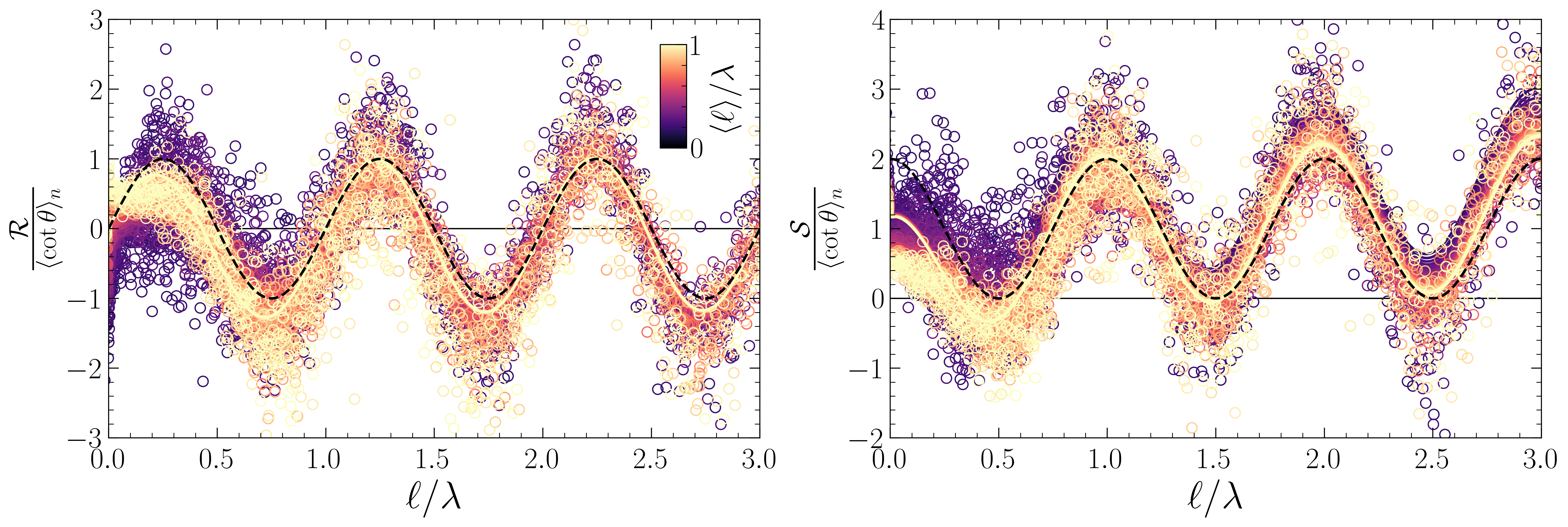}
    \caption{Hoplength amplification rates $\mathcal{R}(k\ell)$ and $\mathcal{S}(k\ell)$ rescaled by $\langle\cot\theta\rangle_n=\int \cot\theta(\ell)P_n(\ell)\dd\ell$ (SI \ref{sec-supp: Wavelength selection}). Circles are DEM measurements for Earth conditions and lines are predictions from the analytical model using geometrical hoplength changes (Fig. \ref{fig-extended data:delta ell}). DEM markers are scaled down by a factor of 1.5 to take into account the missing $\Delta\ell$ contributions. The dashed lines are the approximations $\mathcal{R}\propto \sin k\ell$ and $\mathcal{S}\propto 1+\cos k\ell$ (see SI \ref{sec-supp: Modulation Rate Approximations}). }
    \label{fig-extended data:A}
\end{figure*}

\begin{figure*}[b]
    \centering
    \includegraphics[width=\textwidth]{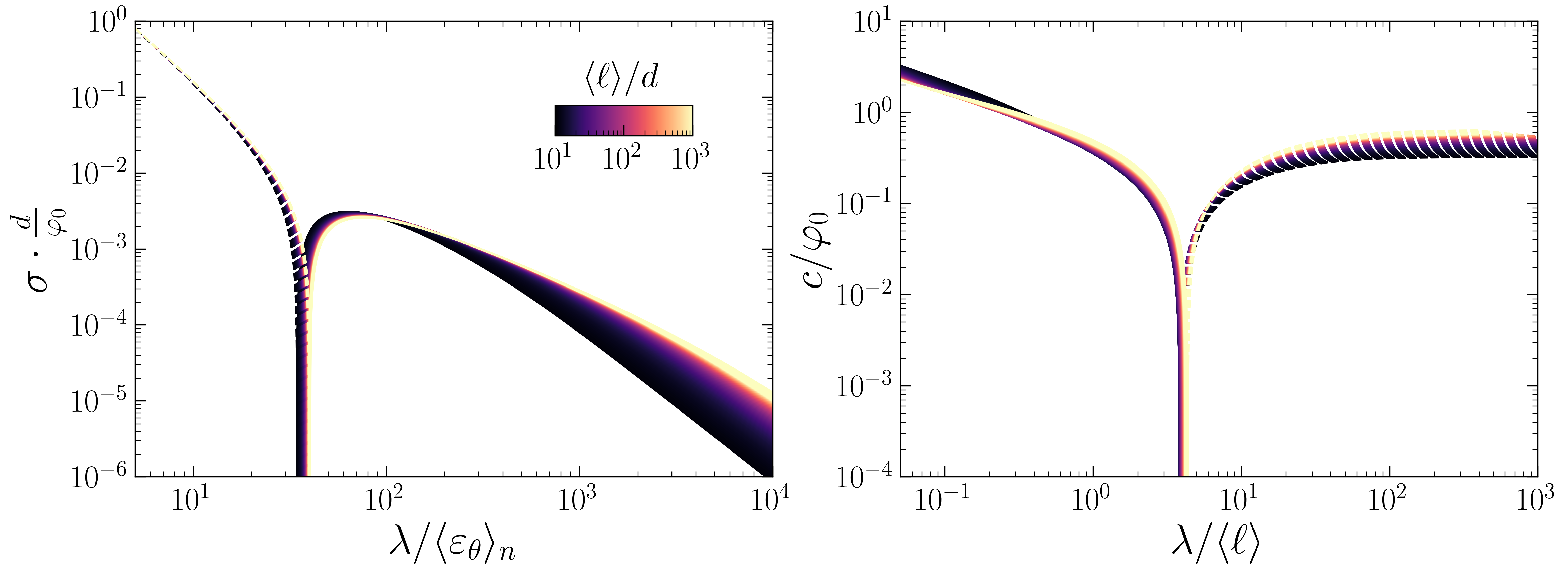}
    \caption{Dispersion relation solutions over a wide range of planetary transport conditions (represented by changing $\langle\ell\rangle$). Positive values are represented by solid lines and negative values by dashed lines. The rescaled growth rate $\sigma$ versus $\lambda$ rescaled by the weighted average lag length $\langle\varepsilon_\theta\rangle_n=\langle\bar\varepsilon\cot\theta\rangle_n/\langle\cot\theta\rangle_n$ shows that the fastest growing wavelength in this model scales with $\langle\varepsilon_\theta\rangle_n$. The rescaled speed $c$ versus $\lambda$ rescaled by average hoplength $\langle\ell\rangle$ shows that the ripple propagation speed in this model is a unique function of $\lambda/\langle\ell\rangle$---and when evaluated at the fastest growing wavelength $\langle\varepsilon_\theta\rangle_n\propto d$ the associated speed is $c/\varphi_0=f(\langle\ell\rangle/d)$. These solutions were derived for $\alpha=0,\ \gamma_n=0.4,\ \gamma_\varepsilon=0.25,\ \gamma_\theta=0.25$ and $\bar\varepsilon_{1d}/d=\cot\theta_{1d}=1$. These parameters represent typical conditions for saltation, $\rho_p/\rho_f\gtrsim 100$ (Table \ref{table-methods: DEM params}). We can see that for $\lambda\sim 200d$ the ripple-antiripple transition ($c=0$) occurs at a critical hoplength $\langle\ell\rangle_c\sim 60d$ consistent with stationary ripples simulations observed for Titian (Fig. \ref{fig4}). To a good approximation the speeds evaluated at the fastest growing wavelength scale as $c/\varphi_0\propto (\langle\ell\rangle/\langle\ell\rangle_c)^{2/5}-1$ for the parameters used here.}
    \label{fig-extended data:disp rel}
\end{figure*}

%

\newpage

\begin{center}
\textbf{\large Supplemental Information: Emergence of wind ripples controlled by mechanics of grain–bed impacts}
\end{center}
\setcounter{equation}{0}
\setcounter{figure}{0}
\setcounter{table}{0}
\setcounter{page}{1}
\makeatletter
\renewcommand{\theequation}{S\arabic{equation}}
\renewcommand{\thefigure}{S\arabic{figure}}

\section{Surface Evolution}\label{sec-supp: Surface Evolution}
\subsection{The Surface}
Our model is based on the fundamental constraint that sand grains are conserved. We thus attempt to evolve the surface $Z(x,t)$ based on this mass conservation constraint, focusing on one spatial dimension $x$ in the direction of the wind (generalization to two spatial dimensions ($x,y$) is straight forward). However, the surface elevation $Z(x,t)$ is not trivially defined \cite{pierce2020joint}. In the simplest case, the surface could be defined by the elevation below which the volume fraction of grains is constant, at the bed packing fraction $\phi_b$. We will call this elevation $Z_b(x,t)$. In practice $\phi_b$ and $Z_b$ are averaged over a finite surface area $\dd x\dd y$ (here $\dd y = 1 d$). For example, if we want to observe elevation changes on the scale of a single grain of diameter $d$, such that $\dd x\dd y = d^2$, then the surface slopes would change drastically and discontinuously over space. Thus we seek to coarse-grain the surface over an area much larger than a single grain but not too large to where we are missing important changes in surface slope. In practice we average the surface over $10d$ in DEM measurements.

The elevation $Z_b(x,t)$ becomes less identifiable as \emph{the surface} as the transport rate $Q$ increase. This is because the volume fraction of grains at a given elevation $\phi(z)$ is no longer a simple step-function between the bed $\phi_b$ and above due to bed dilation from grain-bed impacts. Thus we consider the evolution of the effective surface given transport, $Z(x,t)$, which is defined as
\begin{equation}
    \label{eq-supp: Z = int{phi/phi_b dz} }
    Z(x,t) \equiv \int_{0}^{\infty}\frac{\phi(z,x,t)}{\phi_b}\dd z.
\end{equation} 
 If $M(x,t)$ is the total volume of grains in motion per unit area---proportional to the shear stress $\tau$ above the transport threshold $M\sim \tau -\tau_{\mathrm{th}}$ \cite{pahtz2020unification}---then 
 \begin{equation}
 Z(x,t) = Z_b(x,t) + \frac{M(x,t)}{\phi_b}.
\end{equation}
Note that in steady state transport over a flat bed, $Z(x)$ \emph{must} on average be equal to the elevation of the surface without transport $Z_b(M=0)$ because $Z$ simply represents the elevation if all grains in transport were piled on the surface $Z_b$. Thus, $Z_b(\tau)$ must decrease at the same rate as $M(\tau)$ increases with $\tau-\tau_{\mathrm{th}}$.

\subsection{Mass Conservation}\label{sec-supp: Mass Conservation}
As we have to coarse grain the spatial scale to define the surface elevation $Z(x,t)$, we also coarsen the time scale for surface evolution $\dd t$ to be slow compared to granular time scales---for example, $\sqrt{d/g}$ being (roughly) the time it takes for a grain to fall a distance $d$ under gravity $g$. I.e., if $\dd t= \sqrt{d/g}$ then the evolution of surface height over time $\partial_t Z(x,t)$ would change discontinuously by large amounts. So we take $\dd t$ to be much larger than $\sqrt{d/g}$ but much smaller than the ripple growth time scale $\sigma^{-1}$, defined by the growth rate $\sigma$ which scales with the excess shear stress $\tau-\tau_{\mathrm{th}}$. For reference, the DEM suggests that for Earth-like conditions $\sigma^{-1}\sim \mathcal{O}(10^5\sqrt{d/g})$ for rarefied transport (e.g., for one particle in transport per $50d^2$) and the average travel time for a grain trajectory of hoplength $\ell\sim 100d$ is $\mathcal{O}(10\sqrt{d/g})$.

Given the above considerations, the equation of motion for the surface $Z(x,t)$ is given by the gradient in the total horizontal flux $Q(x,t)$ (normalized by $\phi_b$) defined as the volume of grains crossing a vertical plain per unit width and time:
\begin{equation}\label{eq-supp: dZ/dt = -div{Q}}
    \partial_t Z(x,t) = -\partial_xQ(x,t).
\end{equation}
This is the so-called ``Exner Equation''. Whereas the evolution of the static bed $Z_b(x,t)$ is given by the difference in the vertical fluxes of grains impacting the surface and leaving the surface, $\varphi_\downarrow(x,t)$ and $\varphi_\uparrow(x,t)$ (again, all normalized by $\phi_b$):
\begin{equation}\label{eq-supp: dZ_b/dt = -div{Q}}
    \partial_t Z_b (x,t) = \varphi_\downarrow(x,t)- \varphi_\uparrow(x,t).
\end{equation}

We assume the change in the transport ``activity'' above the static surface over time to be negligible $\partial_t M\simeq 0$ in comparison to $\partial_t Z_b$. This is an argument that the time it takes for transport to adjust is \emph{much} faster than the exceedingly slow time scale evolution of the surface. Thus surface evolution is given by
\begin{equation}\label{eq-supp: dZ/dt = -dQ/dx = Diff varphi}
    \partial_t Z(x,t) = -\partial_xQ(x,t) = \varphi_\downarrow(x,t)- \varphi_\uparrow(x,t).
\end{equation}

\subsection{Linking Impacts to Ejections}\label{sec-supp: link impacts to ejections}

Our model for ripple growth is derived by relating fluxes at each point to upwind fluxes using grain trajectories. The impact flux $\varphi_\downarrow(x_\downarrow)$ is found by relating all grains impacting a location $x_\downarrow$ to the location they left from $x_\uparrow$---linking $\varphi_\downarrow(x_\downarrow)$ to $\varphi_\uparrow(x_\uparrow)$ via the hoplength $\ell=x_\downarrow-x_\uparrow$. This requires information about the distribution of hoplengths, which is generally modulated in response to a modulated surface $Z(x,t)$.

We define $\psi(\ell,x_\downarrow)\dd \ell \dd x_\downarrow$ as the volume of grains (at bed packing fraction $\phi_b$) impacting the surface between $x_\downarrow$ and $x_\downarrow+\dd x_\downarrow$ after hoplengths between $\ell$ and $\ell+\dd \ell$, such that $\varphi_\downarrow(x_\downarrow) \equiv \int\psi(\ell,x_\downarrow)\dd \ell$. The conservation of mass, with reference to Figure \ref{fig2}a,b in the main text, dictates 
\begin{equation}\label{eq-supp:psi=Pphi}
\psi(\ell,x_\downarrow)\dd\ell\dd x_\downarrow = \varphi_\uparrow(x_\uparrow)\dd x_\uparrow P_0(\ell_0)\dd \ell_0,
\end{equation}
which states that the grains impacting at $x_\downarrow$ with \emph{modulated} hoplengths $\ell$ are the same grains that left from $x_\uparrow$ with \emph{unmodulated} hoplengths $\ell_0$ given by the homogeneous probability density $P_0(\ell_0)$ defined on a flat surface.

In order to equate the impact flux density $\psi(\ell,x_\downarrow)$ nonlocally to the upstream vertical flux $\varphi_\uparrow(x_\uparrow)$ given the hoplength distribution $P_0(\ell_0)$ in Equation \eqref{eq-supp:psi=Pphi}, we change coordinates from  $(\ell_0,x_\uparrow)$-space associated with flat bed information to $(\ell,x_\downarrow)$-space which takes into account the changes in the flat bed coordinates due to the modulated surface:

\begin{equation}\label{eq-supp:psi=Pphi transform}
\psi(\ell,x_\downarrow) = \varphi_\uparrow(x_\downarrow-\ell) P_0(\ell-\Delta\ell) | \boldsymbol{\mathcal{J}}_{\ell,x_\downarrow} |,
\end{equation}
where $\Delta\ell(\ell,x_\downarrow)=\ell-\ell_0$ is the small perturbation to the flat bed hoplength $\ell$ from surface modulation effects and $| \boldsymbol{\mathcal{J}}_{\ell,x_\downarrow} |=|1-\partial_\ell\Delta\ell-\partial_{x_\downarrow}\Delta\ell|$ is the resulting Jacobian of the coordinate transformation.

\subsection{Linking Ejections to Impacts}\label{sec-supp: Linking Ejections to Impacts}
Equation \eqref{eq-supp:psi=Pphi transform} puts the impact flux $\varphi_\downarrow$ in terms of the ejection flux $\varphi_\uparrow$ nonlocally using grain trajectories $\ell$. To close the problem we now relate ejections to impacts---which is a more local response. Reiterating the ejection process outlined in the Methods, ejections and rebounds are generally described using the replacement capacity density (per unit lag distance $\varepsilon=x_\uparrow-x_\downarrow$), $n_\varepsilon(\varepsilon|\ell,x_\downarrow)$---which may in general depend on surface geometry and thus may be a function of space. Thus the general rate of ejections and rebounds is given by
\begin{equation}
    \label{eq-supp: varphi_uparrow(x)}
    \varphi_\uparrow(x_\uparrow)=\iint \psi(\ell,x_\uparrow-\varepsilon)n_\varepsilon(\varepsilon|\ell,x_\uparrow-\varepsilon)\dd \varepsilon \dd \ell.
\end{equation}
Where the replacement capacity (total number of ejections/rebounds per impact) is expressed as $n_\uparrow(\ell,x)=\int n_\varepsilon(\varepsilon|\ell,x)\dd\varepsilon$. 
A spatial dependence of $n_\varepsilon$ suggests that there are two modulating effects on the fluxes, one from modulations of trajectories $\Delta\ell$ and the other from modulations of ejections and rebounds caused by impact-ejection interactions with surface geometry. The latter is the main driver for the modulation of surface erosion and the emergence of ion sputter ripples \cite{sigmund1973mechanism,norris2019ion} where ion impacts penetrate into a material and sputter (erode) the surface in a way that greatly depends on surface slope and curvature. Although this is a mechanism similar to the one in which grains are ejected from a sand bed, we find that this effect is in fact small compared to the modulation component coming from $\Delta\ell$---which is reasonable given that grain impacts do not travel deep into the bed surface in comparison to the ion-sputter process \cite{sigmund1973mechanism}. Thus from here on we let the replacement capacity (density) be spatially constant in our analysis, 
\begin{equation}
n_\varepsilon(\varepsilon|\ell,x_\downarrow)\sim n_\varepsilon(\varepsilon|\ell).
\end{equation}
A more general derivation of the model with spatially explicit ejection dependence is straight forward.

\section{Linear Instability}\label{sec-supp: linear instability}
\subsection{Linear Regime}\label{sec-supp: linear regime}
In this study we are mainly concerned with the \emph{linear} instability of ripple formation from the uniform flat bed fixed point $Z_*(x)\equiv0$. In the linear regime we know that perturbations to the surface $Z(x,t)$ must evolve as $\partial_t Z\sim \Omega Z$ where the growth rate $\Omega$ is a linear operator that determines how perturbation amplitudes grow vertically and propagate in space---i.e., $\Omega$ consists of local operations $\partial_x, \partial_{xx}, \partial_{xxx}...$ and much more complicated spatial dependencies because the problem is generally nonlocal. The criterion for the linear regime is $|\partial_xZ|\ll 1$ for all $x$---for an \emph{average} surface elevation over $\dd x$ bigger than a grain diameter $d$. Linearity allows us to explicitly expand space and time dependent quantities into a zeroth order component associated with flat-bed/steady-state transport and a first order component that takes into account small changes in proportion with $Z(x,t)$. Thus for example we let $\varphi(x,t)=\varphi_0 + \varphi_1(x,t)$ and $  \psi(\ell,x,t)=\psi_0(\ell)+\psi_1(\ell,x,t)$ . This allows the growth rate operator to be written explicitly as
\begin{equation}
\Omega = \frac{\varphi_{\downarrow 1}(x,t) -\varphi_{\uparrow 1}(x,t)}{Z(x,t)}
\end{equation}
where we have used $\varphi_{\uparrow 0}=\varphi_{\downarrow 0}=\varphi_0$, which defines steady state on a flatbed. This shows that \emph{only} first order perturbations to the fluxes are responsible for ripple growth. Because of this we will drop the subscript's-1 for all first order (spatially dependent) expressions below and explicitly denote flat bed (spatially constant) quantities with subscript's-0.

Imposing linearity, the first order perturbation to the impact flux density $\psi(\ell,x)$ in Equation \eqref{eq-supp:psi=Pphi transform} simplifies to
\begin{align}\label{eq-supp: psi1(l,x)}
\psi(\ell,x)& = \varphi_{\uparrow }(x-\ell)P_0(\ell)- \varphi_0\partial_{\ell,x} \Big(\Delta\ell(\ell,x) P_0(\ell)\Big) 
\end{align}
where $\partial_{\ell,x}=\partial_\ell +\partial_x$. This is the same as Methods Equation \eqref{eq-methods: psi(l,x)} and integrates to give the total impact flux $\varphi_\downarrow$ in Methods Equation \eqref{eq-methods: varphi_down(x)}.

With Equation \eqref{eq-supp: psi1(l,x)} (and Methods Eq. \ref{eq-methods: varphi_down(x)}), the surface evolution Equation \eqref{eq-supp: dZ/dt = -dQ/dx = Diff varphi} can now be written to first order as
\begin{eqnarray}
\label{eq-supp: dZdt= blah + dl-correction}
\partial_t Z(x,t)  &=& \int \varphi_{\uparrow }(x-\ell,t)P_0(\ell) \dd \ell - \varphi_{\uparrow }(x,t) -\varphi_0\partial_x\langle\Delta\ell(x,t)\rangle.
\end{eqnarray}
 We see that the first term assumes no explicit changes to grain hoplengths---as has been commonly used in past impact ripple research \cite{anderson1987theoretical,csahok2000dynamics}---and the final term is a result of the extra flux associated with small changes in grain trajectories in the presence of the modulated surface.

Alternatively, using the definition of $\varphi_\uparrow$ from Equation \eqref{eq-supp: varphi_uparrow(x)} the evolution of the surface may be described in terms of impacts and resulting bed response from ejections:
\begin{equation}
\label{eq-supp: dZdt = int craters}
\partial_t Z(x,t)  = \iint \psi(\ell,x-\varepsilon,t)\Big[\delta(\varepsilon)-n_\varepsilon(\varepsilon|\ell) \Big] \dd\varepsilon\dd\ell,
\end{equation}
where the net number of grains exchanged with the bed $\delta(\varepsilon)-n_\varepsilon(\varepsilon|\ell)$ is analogous to the ``crater function'' used in describing the growth of ion-sputter ripples \cite{norris2019ion}.

\subsection{Growth and Modulation Rates}\label{sec-supp: Growth and Modulation Rates}
We can greatly simply the analytical expressions above by independently analyzing perturbations of a given (Fourier) mode $k=2\pi/\lambda$,
\begin{equation}
Z_k(x,t)=\hat{Z}(k,t)e^{ikx}
\end{equation}
where the sum over all $k$ recovers the surface $Z(x,t)=(2\pi)^{-1}\int Z_k(x,t)\dd k$, keeping the real part only. In the linear growth regime, $k|\hat{Z}|\ll 1$, the complex amplitude evolves according to $\hat{Z}(t) \propto e^{\Omega t}$ where the operator $\Omega(k)$ at a given mode $k$ turns into a complex growth rate, 
\[\Omega(k) =\sigma(k) + i\omega(k),\]
with $\sigma$ being the real growth rate of the magnitude $|\hat{Z}|$ and $c=-\omega/k$ is the propagation speed of the perturbed mode $k$.

At mode $k$ all other spatially explicit (first order) quantities can be cast as Fourier amplitudes proportional to $\hat{Z}$. We thus let 
\begin{equation}\label{eq-supp: def of F mod rates}
\hat\varphi_{\downarrow} = \varphi_0\mathcal{F}_{\downarrow}k\hat{Z} \quad \text{and}\quad \hat\varphi_{\uparrow} = \varphi_0\mathcal{F}_{\uparrow}k\hat{Z}
\end{equation}
which define the vertical flux modulation rates $\mathcal{F}_{\uparrow(\downarrow)}$ as complex numbers whose magnitude gives the size of the flux perturbation and whose phase locates the modulated flux relative to the surface phase. For example, $\arg \mathcal{F}_{\downarrow}=\pi/2$ says that the perturbation to the impact flux $\hat{\varphi}_\downarrow$ is located at the upwind inflection point---i.e., $\pi/2$ out of phase with $\hat{Z}$. For the impact flux density and hoplength modulations we let
\begin{equation}
\hat{\psi}(\ell)=\psi_0(\ell)\mathcal{A}(\ell)k\hat{Z}\quad \text{and}\quad \Delta\hat{\ell}(\ell)=\mathcal{L}(\ell)\hat{Z}
\end{equation}
which defines the complex modulation rates $\mathcal{A}(\ell)$ and $\mathcal{L}(\ell)$, which encode the same information as $\mathcal{F}_{\uparrow(\downarrow)}$ but on a per hoplength basis. For example, $\mathcal{A(\lambda)}$ gives the amplification of the impact flux for grains with hoplength $\ell=\lambda$. We note that for geometrical hoplength changes (see Methods Eq. \ref{eq-methods: Dl}), the hoplength modulation rate is given by
\begin{equation}\label{eq-supp: L}
    \mathcal{L}(\ell)=\cot\theta(\ell)\left(e^{-ik\ell}-1\right),
\end{equation}
where we used the identity $Z(x_\uparrow)=Z(x_\downarrow-\ell)=\hat{Z}e^{ikx_\downarrow}e^{-ik\ell}$.

Given these definitions we can simplify the surface evolution Equations \eqref{eq-supp: dZ/dt = -dQ/dx = Diff varphi} and \eqref{eq-supp: dZdt= blah + dl-correction} to find the so-called ``dispersion relation'' for the growth rate $\Omega(k)$:
\begin{align}\label{eq-supp: Omega(k)}
    \frac{\Omega}{k\varphi_0} &= \mathcal{F}_\downarrow-\mathcal{F}_\uparrow =(\widetilde{P}_0-1)\mathcal{F}_\uparrow - i\langle\mathcal{L}\rangle
\end{align}
where $\widetilde{P}_0=\int P_0(\ell)e^{-ik\ell}\dd\ell$. The ascending flux modulation rate is defined by Equation \eqref{eq-supp: varphi_uparrow(x)} as 
\begin{equation}
\label{eq-supp: Fup=int{A...}}
\mathcal{F}_\uparrow= \int \tilde{n}_{\varepsilon }(\ell)\mathcal{A}(\ell)P_0(\ell)\dd\ell,
\end{equation}
where $\tilde{n}_{\varepsilon }(\ell)=\int n_{\varepsilon}(\varepsilon|\ell) e^{-ik\varepsilon}\dd\varepsilon$. Thus we are also able to write the dispersion relation in terms of the amplification of impacting trajectories and the resulting bed response (Eq. \ref{eq-supp: dZdt = int craters}):

\begin{equation}
    \label{eq-supp: Omega = int{A...}}
    \frac{\Omega}{k\varphi_0} = \int(1-\tilde{n}_\varepsilon(\ell)) \mathcal{A}(\ell)P_0(\ell)\dd\ell.
\end{equation}

From the derivation of $\psi(\ell,x)$ in Equation \eqref{eq-supp: psi1(l,x)}, the hoplength amplification rate is
\begin{align}
    \label{eq-supp:A}
    \mathcal{A} &= \mathcal{F}_\uparrow e^{-ik\ell} - k^{-1}\Big\{ \mathcal{L}'+ \big(P_0'/P_0 +ik\big)\mathcal{L}\Big\}, 
\end{align}
where $(\cdot)'=\partial_\ell(\cdot).$ Thus we obtain a closed-form solution to $\Omega(k)$ by solving for $\mathcal{F}_\uparrow$ using Equations \eqref{eq-supp: Fup=int{A...}} and \eqref{eq-supp:A} to obtain
\begin{align}\label{eq-supp: Fup}
\rho_n\mathcal{F}_\uparrow & = -i\int \tilde{n}_{\varepsilon} \mathcal{L}P_0\dd\ell +k^{-1}\int \tilde{n}_{\varepsilon }'\mathcal{L}P_0\dd\ell,
\end{align}
where $\rho_n=1-\int \tilde{n}_{\varepsilon} P_0 e^{-ik\ell}\dd\ell$ is a factor associated with the condition that the vertical flux is in modulated equilibrium (Fig. \ref{fig2}b). 

\subsection{Local and Nonlocal Regimes}\label{sec-supp: local nonlocal}

\begin{figure}[t]
    \centering
    \includegraphics[width=.5\textwidth]{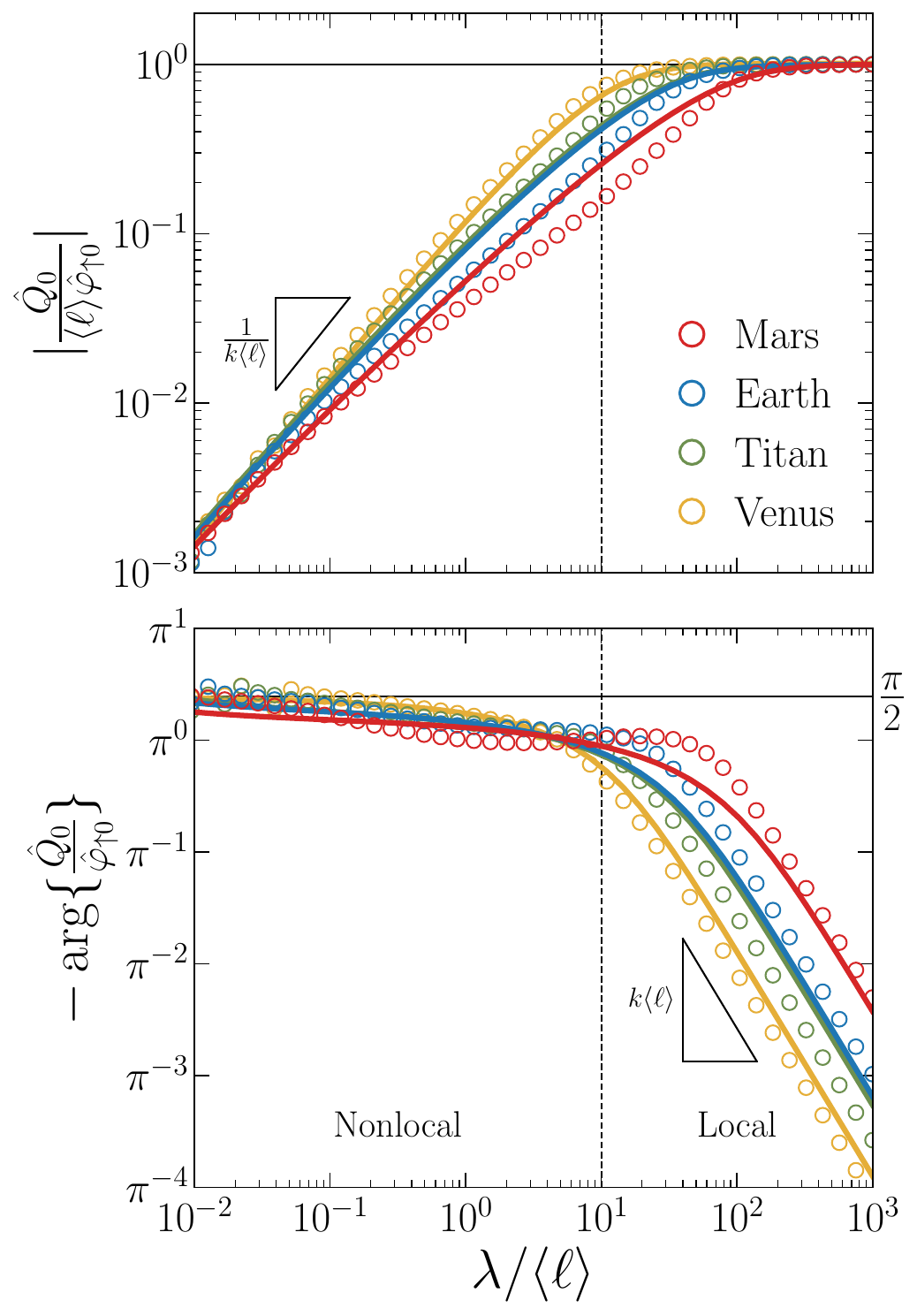}
    \caption{Magnitude and phase relationships for the modulated flux $\hat{Q}_0$ responding to a sinusoidal ejection flux $\hat{\varphi}_{\uparrow 0}$ at mode $k=2\pi/\lambda$ on a flatbed. Markers are derived from DEM measurements of $P_0(\ell)$ for Mars, Earth, Titan and Venus conditions (see Fig. \ref{fig3}) and the solid lines are using power-law approximations of $P_0(\ell)$ with exponential cutoffs (Methods; Table \ref{table-methods: DEM params}). }
    \label{fig-supp: 1-P0_}
\end{figure}

It is useful to analyze the linear instability dynamics in terms of `local' and `nonlocal' flux response, which arise from the relationship between the perturbation size $\lambda$ and the average hoplength $\langle\ell\rangle$. In this system these regimes are identified via the function $1-\widetilde{P}_0$, which represents the modulated relationship between the flux modulation $\hat{Q}_{0}$ in response to a sinusoidal modulation of the ejection flux $\hat{\varphi}_{\uparrow 0}$ on a flatbed (i.e., no $\Delta\ell$ contributions):
\begin{equation}\label{eq-supp: 1-P0_}
\frac{1-\widetilde{P}_0}{ik} \equiv \frac{\hat{Q}_0}{\hat{\varphi}_{\uparrow 0}}\sim
    \begin{dcases}
         \langle\ell\rangle e^{-\gamma ik\langle\ell\rangle}, & \lambda/\langle\ell\rangle\gg1 \\
        \lambda e^{-i\pi/2}, &  \lambda/\langle\ell\rangle\lesssim1.\\
    \end{dcases}
\end{equation}
In the limit of long wavelength modulations relative to $\langle\ell\rangle$, the flux response is local and simply reduces to the identity $\hat{Q}_0\sim\langle\ell\rangle\hat{\varphi}_{\uparrow 0}$ with a downwind lag of $\hat{Q}_0$ relative to $\hat{\varphi}_{\uparrow 0}$ proportional to $\langle\ell\rangle$ (Fig. \ref{fig-supp: 1-P0_})---with $\gamma$ being the proportionality constant in Eq. \eqref{eq-supp: 1-P0_}. For wavelengths commensurate with $\langle\ell\rangle$, the phase relation between $\hat{Q}_0$ relative to $\hat{\varphi}_{\uparrow 0}$ grows because of the nonlocal contributions from the fluxes upwind and asymptotically approaches a relative phase shift of $\pi/2$. This asymptotic phase shift behavior is purely due to the power-law-like shape of $P_0(\ell)$ which provides exceedingly large local contributions ($\ell/\lambda\lesssim1$) to $\hat{Q}_0$ that dominate in the long-wavelength regime, but are counteracted by the by the increasing contributions from trajectories originating on other parts of a ripple ($\ell/\lambda\gtrsim1$) as wavelength decreases relative to $\langle\ell\rangle$. And in the nonlocal limit the flux ratio magnitude $|\hat{Q}_0/\hat{\varphi}_{\uparrow 0}|\sim \lambda$ because the nonlocal flux contributions from upwind scale with $\lambda$, almost irrespective of $P_0(\ell)$ (for small enough $\lambda/\langle\ell\rangle$).

\subsection{Salton-Repton Model}\label{sec-supp: Salton-Repton Model}

\begin{figure*}[t]
    \centering
    \includegraphics[width=\textwidth]{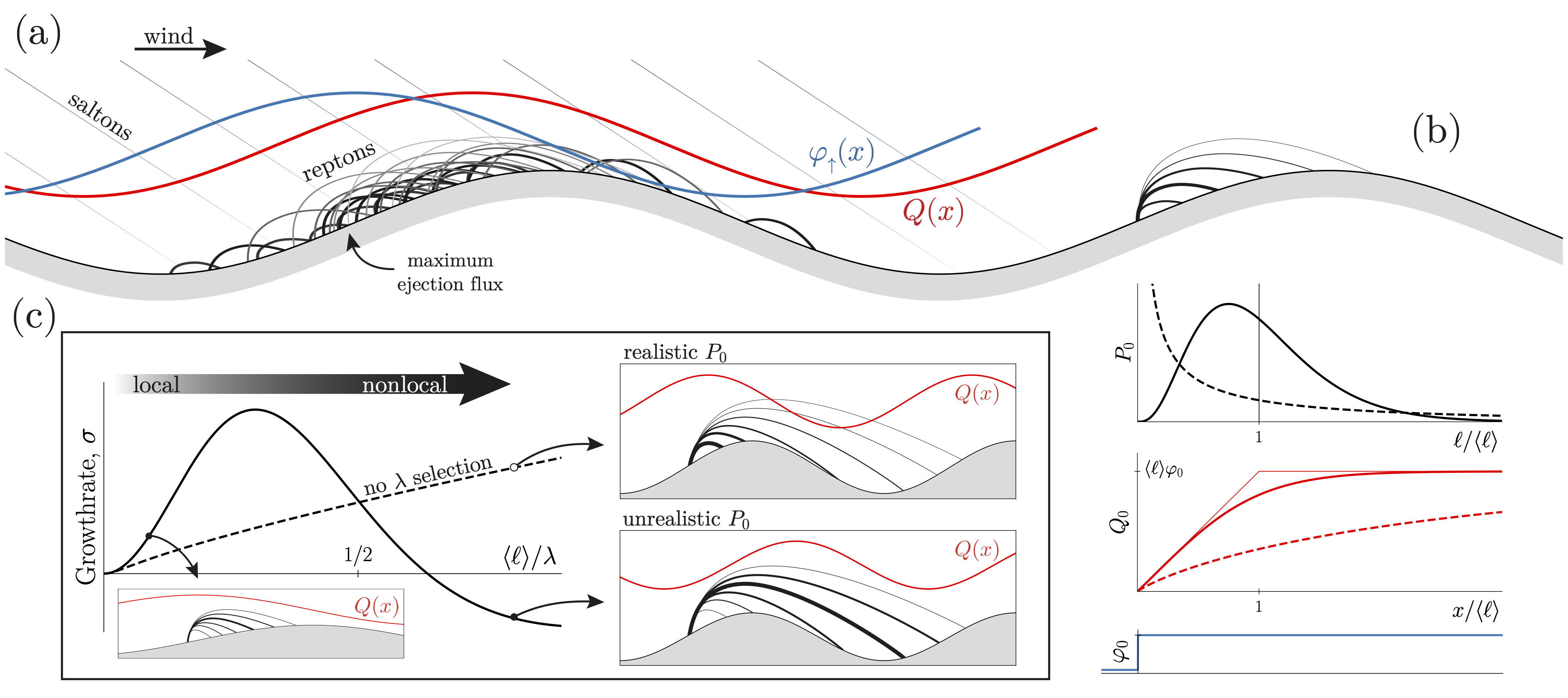}
    \caption{(a) Schematic of the widely used salton-repton model for ripple formation. A uniform salton rain is focused on upwind faces and causes ejections of reptons. Thus the maximum in the ejection flux $\varphi_\uparrow(x)$ (blue) is located at the upwind inflection points and the horizontal flux $Q(x)$ (red) is shifted downwind. Note that for the schematic drawn in (a), the flux $Q(x)$ is converging ($\partial_xQ < 0$) at the crest causing this particular ripple wavelength to grow. Section (b) on the right shows a set of grain trajectories being transported from the upwind inflection point. The thickness of each trajectory is given by the number of grains, or probability $P_0(\ell)$ with that given hoplength---taken to be Gaussian-like. The $P_0$ plot shows Gaussian-like (solid line), power-law like (dashed line) and delta (thin vertical line) distributions. The $Q_0(x)$ plot shows transient relaxation effects for a sudden step in the ejection flux to $\varphi_0$ at $x = 0$ (bottom plot, blue line) for each of the three distributions above it. Panel (c) shows the growth rates versus rescaled wavenumber $\langle\ell\rangle/\lambda$ for the Gaussian-like $P_0$ (solid line; ``unrealistic") and power-law like $P_0$ (dashed line; ``realistic"). We are also showing in the insets of (c) that the ratio $\langle\ell\rangle/\lambda$ separates the dynamics from local ($\langle\ell\rangle/\lambda \ll 1$), where the form of $P_0$ has little importance, to nonlocal ($\langle\ell\rangle/\lambda \gtrsim 1$) where the form of $P_0$ is crucial. Red lines in the insets represent the flux $Q(x)$.}
    \label{fig-supp: salton-repton model}
\end{figure*}
The seminal salton-repton approach to understanding wind ripple formation was first formalized by Anderson \cite{anderson1987theoretical} and then later iterations of this model were proposed, all with similar concepts and predictions \cite{prigozhin1999nonlinear,csahok2000dynamics, duran2014direct}---notably the prediction that ripple wavelengths are selected by a transport length scale, the average hoplength or repton length. In this framework, a uniform salton rain bombards the surface and causes ejections of reptons. Because the salton rain is uniform (unmodulated) and the impact-ejection lag length $\varepsilon$ is ignored, the maximum in the ejection flux of reptons is located at the upwind inflection points where slopes are greatest. In our framework, this translates to $\varphi_\uparrow\propto \partial_x Z$  or $\mathcal{F}_\uparrow\propto i$ (Eq. \ref{eq-supp: def of F mod rates}) and thus 
\begin{equation}\label{eq-supp: Omega salt-rep}
    \Omega_{\text{salt-rep}} \propto ik(\widetilde{P}_0-1)\propto k^2 \frac{\hat{Q}_0}{\hat{\varphi}_{\uparrow 0}}
\end{equation}
where $P_0(\ell)$ can now effectively be interpreted and the distribution of repton hoplengths. In the salton-repton model the horizontal flux $Q(x)$ is shifted downwind of the ejection flux $\varphi_\uparrow(x)$ by a distance depending only on the homogeneous probability distribution of hoplengths $P_0(\ell)$ (Fig. \ref{fig-supp: salton-repton model}a,b)---and doesn't take into account effects from changes in hoplengths $\Delta\ell$ or the impact-ejection lag $\varepsilon$ (Fig. \ref{fig2}). And because of the power-law like structure of $P_0$ (Fig. \ref{fig3}), the salton-repton model predicts that the growth rate $\sigma_{\text{salt-rep}}(k)>0$ for all $k$ because the phase shift $\arg\{\hat{Q}_0/\hat{\varphi}_{\uparrow 0}\}$ doesn't allow for the flux maximum to pass the crest for $\hat{\varphi}_{\uparrow}=\hat{\varphi}_{\uparrow 0}$ maximized at the upwind infection point, $\pi/2$ out of phase (Fig. \ref{fig-supp: 1-P0_}). In other words, the salton-repton model precludes erosion of the ripple crest for all $k$ (see Figures \ref{fig4}, \ref{fig-supp: 1-P0_} and \ref{fig-supp: salton-repton model}). 

It is often assumed that $P_0(\ell)$ is Gaussian-like, which is unphysical because ejections arise from a broad continuum of impacting grain energies (Fig. \ref{fig3}). Another common approach is to assume a single (reptating) grain trajectory $\ell=\langle\ell\rangle$ such that $P_0\sim \delta(\ell - \langle\ell\rangle)$. Figure \ref{fig-supp: salton-repton model} shows the consequences of these unrealistic Gaussian-like and delta hoplength distributions when compared to a more realistic power-law like $P_0(\ell)$. Figure \ref{fig-supp: salton-repton model}b shows the transient effects of a sudden step in the flatbed ejection flux $\varphi_0$ at $x = 0$ on the resulting relaxation of the horizontal flux $Q_0(x)$. Here the flatbed horizontal flux $Q_0(x)$ is the volume of grains ejected at $x - \ell$ that exceed the travel distance $\ell$ \cite{furbish2012probabilistic}. Thus $Q_0(x) = \varphi_0 \int_0^x R(\ell)\dd \ell$ relaxes to its saturated state $\langle\ell\rangle\varphi_0$ depending on the shape of $R(\ell) = \int^\ell P_0\dd\ell$. For Gaussian/delta-like distributions this relaxation occurs quickly compared to $\langle\ell\rangle$ whereas for the power-law like distribution the relaxation is much slower---saturating at $x\gg \langle\ell\rangle$. This means that the effects of perturbations to the vertical flux $\varphi_\uparrow$ can be felt far downwind for power-law like $P_0(\ell).$

Panel (c) of Figure \ref{fig-supp: salton-repton model} shows how Gaussian/delta-like $P_0$ effects ripple growth. The growth rates for the salton-repton model for the Gaussian-like $P_0$ gives rise to a fastest growing wavelength that scales with the average hoplength $\lambda\sim \langle\ell\rangle$ and approaches $\lambda\rightarrow 4\langle\ell\rangle$ as $P_0 \rightarrow \delta(\ell - \langle\ell\rangle)$ (Eq. \ref{eq-supp: Omega salt-rep}). However for the more realistic power-law like $P_0$ there is no maxima and thus no wavelength selection (the smallest wavelength grows the fastest). Also highlighted in Figure \ref{fig-supp: salton-repton model}c (insets) is the fact that in the local regime $\langle\ell\rangle/\lambda \ll 1$ the form of $P_0$ has little importance, however in the nonlocal regime $\langle\ell\rangle/\lambda \gtrsim 1$ the form of $P_0$ is crucial: For $\langle\ell\rangle/\lambda \ll 1$ grains are redistributed locally on the ripple face---thus the flux $Q$ is lagged from $\varphi_\uparrow$ by a small amount (set by $\langle\ell\rangle$) relative to $\lambda$. However, in the $\langle\ell\rangle/\lambda \gtrsim 1$ regime, $Q(x)$ at any spot $x$ is nonlocally related to $\varphi_\uparrow(x)$ far upwind. Thus for the unrealistic $P_0(\ell)$, the peak of the Gaussian-like density of grains tends to cause a large lag between $\varphi_\uparrow$ and $Q$, shifting the maximum of $Q$ past the crest and leading to erosion of the crest and stabilization of small wavelengths. This can also be viewed as grains filling up the downwind troughs. However for the realistic $P_0(\ell)$, the flux relaxes so slowly to changes in $\varphi_\uparrow$ (Fig. \ref{fig-supp: salton-repton model}b), because the grains are weighted heavily at $\ell = 0$, that the flux maximum never passes the crest, even as $\lambda\rightarrow0$. Therefore, the salton-repton model, without inclusion of an impact-ejection lag, does not include a stabilizing mechanism and therefore cannot explain wavelength selection for impact ripples.

\subsection{Modulation Rate Approximations}\label{sec-supp: Modulation Rate Approximations}

To gain insight into the behavior of the flux modulations $\mathcal{F}_\uparrow$ and $\mathcal{A}(\ell)$ we look at the dynamics around the marginal stability point $\Omega(k_*)=0$ where the bed perturbation $k_*$ does not grow or propagate. This occurs when the magnitude and phase of the vertical fluxes match, $\mathcal{F}_\downarrow=\mathcal{F}_\uparrow$, which occurs for wavelengths of order $\langle\ell\rangle$ where the hoplength amplification effects for $\mathcal{A}(\ell\sim \lambda)$ are significant. Note that $k_*$ is of order (around 2 times) the fastest growing mode and thus the dynamics around $k_*$ tell us about the dynamics that give rise to the ripples that emerge from a flat bed. 

Using Equation \eqref{eq-supp: Omega(k)} we conclude that the marginal vertical flux modulations are $\mathcal{F}_*= i\langle\mathcal{L}\rangle/(\widetilde{P}_0-1)$. If we approximate the hoplength modulation as being purely geometric (Eq. \eqref{eq-supp: L}) we can approximate the flux modulation around the marginal stability mode as \[\mathcal{F}_\uparrow\sim \mathcal{F}_* + \delta\mathcal{F}_\uparrow \sim i\langle \cot\theta\rangle + \delta\mathcal{F}_\uparrow
\] where for illustration we take $\mathcal{F}_*=i\langle\cot\theta\rangle$ and absorb the error in $\delta\mathcal{F}_\uparrow.$
We can see that the correction to the vertical flux $\delta\mathcal{F}_\uparrow$ is responsible for ripple growth around $k_*$: \[\frac{\Omega}{k\varphi_0} = (\widetilde{P}_0-1)(\mathcal{F}_\uparrow-\mathcal{F}_*)= (\widetilde{P}_0-1)\delta\mathcal{F}_\uparrow.\] 
The point here is that around the mode $k_*$ (of order the fastest growing mode), the hoplength amplification forcing is strong and thus tends to maximize the vertical flux $\varphi_\uparrow$ at the upwind inflection point (to 0-th order). We can see the amplified forcing around $k_*$ by observing $\mathcal{A}_*(\ell|\mathcal{F}_*)$ using equation \eqref{eq-supp:A} for $\ell/\lambda\gtrsim 1$:
\[\frac{\mathcal{A}_*}{\langle\cot\theta\rangle}\sim i\left(e^{-ik\ell}+1\right)\]
where we have ignored the small contributions from $-k^{-1}P_0'/P_0\sim (k\ell)^{-1}$. This expression highlights that $\mathcal{A}=\mathcal{A}_* + \delta\mathcal{A}$ is approximately a unique function of $\ell/\lambda$, thus showing that the forcing from trajectories $\ell\sim \lambda$ is approximately scale invariant, $\mathcal{A}(\ell\sim \lambda)\sim
\mathcal{A}_*(\ell/\lambda)$. The forcing amplification is seen via $\mathcal{A}_*(\ell=\lambda)\simeq 2i\langle\cot\theta\rangle$ which is maximized around the upwind inflection point ($\pi/2$ out of phase; $i=e^{i\pi/2}$). This scale-invariant forcing is observed in DEM measurements of $\mathcal{A}(\ell)$ seen in Methods Figure \ref{fig-extended data:A}.

\begin{figure}[t]
    \centering
    \includegraphics[width=.5\textwidth]{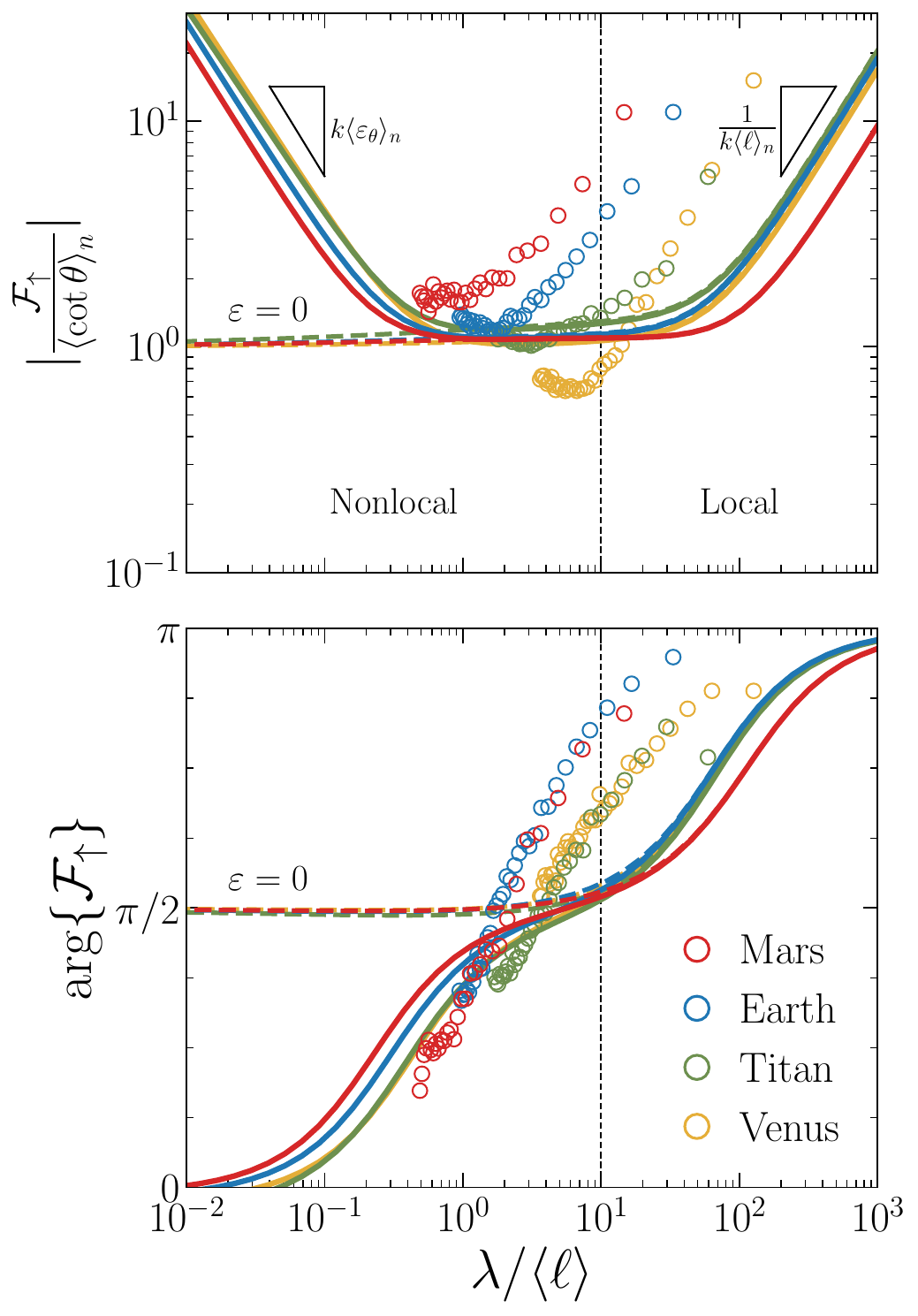}
    \caption{Magnitude and phase of the ejection flux modulation rate $\mathcal{F}_\uparrow=\hat{\varphi_\uparrow}/(\varphi_0k\hat{Z})$. Symbols are measurements from DEM simulations and lines are approximations using geometrical hoplength changes $\mathcal{L}=\cot\theta(\ell)(e^{-ik\ell}-1)$ in Equation \ref{eq-supp: Fup}. The dashed lines show the large-scale/0-th order modulation $\mathcal{F}_\uparrow(\varepsilon=0)\simeq f_\uparrow$ (Eq. \ref{eq-supp: Fup approx}). The disagreement between measured and approximated $\mathcal{F}_\uparrow$ is likely due to measurement error and missing contributions to $\mathcal{L}$ (see Methods).}
    \label{fig-supp: Fup}
\end{figure}

\subsection{Multiple-scales: Impact-Ejection Length Selects the Wavelength}\label{sec-supp: Wavelength selection}

Because we are only interested in the regime where $k\varepsilon\ll 1$ (namely, $\lambda\gg d$) we can approximate $\tilde{n}_{\varepsilon }(\ell)=\int n_\varepsilon(\varepsilon|\ell)e^
{-ik\varepsilon}\dd\varepsilon\sim n_{\uparrow }(\ell)(1-ik\bar\varepsilon(\ell))$. The modulation rate $\mathcal{F}_\uparrow$ can then be expanded into a large-scale/0-th order modulation, which reflects the $\Delta\ell$ modulations of the impact forcing, and a small-scale/1-st order modulation contribution reflecting the downwind shift in ejections from $\varepsilon$. We can approximate the modulation rate generally as 
\begin{equation}\label{eq-supp: Fup approx}
\mathcal{F}_{\uparrow}  \simeq f_\uparrow\big(1-ik\langle \varepsilon_{\Delta\ell}\rangle_n\big)\simeq f_\uparrow e^{-ik\langle \varepsilon_{\Delta\ell}\rangle_n}
\end{equation}
where $f_\uparrow = \rho_n^{-1}\big(-i\langle\mathcal{L}\rangle_n +k^{-1}\langle n_{\uparrow}'\mathcal{L}\rangle\big)$ is the large-scale/0-th order modulation which is an approximate reflection of the forcing $f_\uparrow\sim\int n_\uparrow(\ell) \mathcal{A}(\ell|f_\uparrow)P_0(\ell)\dd\ell$---or put another way, $f_\uparrow\sim \mathcal{F}_\uparrow(\varepsilon=0)$. Taking into account the nonlocal impact-ejection lag $\varepsilon$, $f_\uparrow$ then becomes phase-shifted downwind by the small length scale $\langle \varepsilon_{\Delta\ell}\rangle_n=\langle\bar\varepsilon\mathcal{L}\rangle_n/\langle\mathcal{L}\rangle_n$ which is the average of the lag $\bar\varepsilon$ weighted by the changes in hoplengths over the distribution $P_n=n_{\uparrow } P_0$ where we define $\langle\cdot\rangle_n=\int(\cdot )P_n\dd\ell$. 

\subsubsection{Local Limit: $\lambda\gg \langle\ell\rangle$}

In the local regime ($\lambda\gg\langle\ell\rangle$) the geometrical hoplength modulation becomes determined by the local slope such that $\mathcal{L}\sim -ik\ell\cot\theta$ and effects from $\varepsilon$ are minor. Therefore the vertical flux modulation in this limit behaves as (Fig. \ref{fig-supp: Fup})
\[\lim_{k\langle\ell\rangle\rightarrow0}\ \frac{\mathcal{F}_\uparrow}{\langle \cot\theta_\ell \rangle_n}\sim i - k^{-1}\langle n_{\uparrow \ell\theta}' \rangle \]
where $\langle \cot\theta_\ell \rangle_n=\langle \ell \cot\theta \rangle_n/\langle\ell\rangle_n$ and $\langle  n_{\uparrow \ell\theta}' \rangle=\langle  n_{\uparrow}' \ell \cot\theta \rangle/\langle\ell \cot\theta\rangle_n$. This shows that the vertical flux tends to increase with the slope $\mathcal{F}_\uparrow\sim i$ as in the idealized salton-repton model, but for low slopes (large $\lambda$) the vertical flux modulation also tends to be maximized in the ripples trough do to the gradients in ejections $n_\uparrow'(\ell)$---that become important when the impacts are shorted or lengthened (impact energy lost or gained) from local slope effects.

Because the effects of $\varepsilon$ are vanishing in the limit as $\langle\ell\rangle/\lambda \rightarrow 0$, ripples grow and propagate according to 

\begin{eqnarray}\label{eq: sig c for fup*exp(-ik epsilon)}
&&\lim_{k\langle\ell\rangle\rightarrow0}\ \frac{\sigma}{k^2Q_0} \sim \langle\cot\theta_\ell\rangle_n-\langle\cot\theta_\ell\rangle >0 \notag\\
&&\lim_{k\langle\ell\rangle\rightarrow0}\ \frac{c}{\varphi_0} \sim -\frac{\langle\ell\rangle}{\langle\ell\rangle_n}\langle n_{\uparrow}'\ell\cot\theta \rangle <0, 
\end{eqnarray}
showing that flat bed ripple growth for large wavelengths occurs through antidiffusion ($\partial_tZ\sim -Q_0\partial_{xx}Z$) at a rate given by the flux $Q_0=\langle\ell\rangle\varphi_0$ and that the speeds in this regime are purely negative (antiripples).

\subsubsection{Nonlocal Limit: $\lambda\ll \langle\ell\rangle$}

 For wavelengths around the fastest growing wavelength, the flux response is nonlocal ($\lambda\lesssim\langle \ell\rangle$; Fig. \ref{fig4}, \ref{fig-supp: salton-repton model}) and thus the impact flux amplification/focusing on the upwind inflection point dominates the forcing. This renders the large scale modulation of the ejection flux (from $\Delta\ell$ effects) maximized around the inflection point, $f_\uparrow\sim i\langle\cot\theta\rangle_n$ (Fig. \ref{fig-supp: Fup}). Thus, in the far nonlocal regime around the maximum of $\sigma$ we have

\begin{eqnarray}\label{eq: Omega_Dl(k) approx kl>>0}
    &&\lim_{k\langle\ell\rangle\rightarrow\infty}\frac{\sigma}{k\varphi_0\langle\cot\theta\rangle_n} \sim \rho_{\Im} + \frac{\langle\mathcal{L}\rangle_{\Im}}{\langle\cot\theta\rangle_n}- \rho_{\Re}k\langle \varepsilon_{\theta}\rangle_n \notag\\
    &&\lim_{k\langle\ell\rangle\rightarrow\infty}\frac{c}{\varphi_0 \langle\cot\theta\rangle_n} \sim  \rho_{\Re} + \frac{\langle\mathcal{L}\rangle_{\Re}}{\langle\cot\theta\rangle_n}+ \rho_{\Im}k\langle \varepsilon_{\theta}\rangle_n 
\end{eqnarray}
where $\rho_{\Re}=1-\widetilde{P}_0|_{\Re}<1$ and $\rho_{\Im}=-\widetilde{P}_0|_{\Im}\ll1$, are both positive and weak functions of $k\langle\ell\rangle$ (for $\lambda\ll\langle\ell\rangle$), and $\langle \varepsilon_{\theta}\rangle_n=\langle\bar\varepsilon \cot\theta\rangle_n/\langle\cot\theta\rangle_n$. This recovers the empirical scaling found by Duran et al. (2014) \cite{duran2014direct} with $\sigma\propto Ak - Bk^2\langle \varepsilon_{\theta}\rangle_n$. Here $A=\langle\cot\theta\rangle_n\rho_{\Im} + \langle\mathcal{L}\rangle_{\Im}$ and $B=\langle\cot\theta\rangle_n\rho_{\Re}$ are weak functions of $k$ in the nonlocal regime. This shows that we can approximate that the fastest growing wavelength scales roughly with the average impact-ejection lag length, $\lambda \sim \frac{B}{A}\langle \varepsilon_{\theta}\rangle_n$. Inserting this scaling into $c(k)$ we find the speed scales with transport dynamics through $\varphi_0$ and $\langle\ell\rangle$, $c\sim \varphi_0 f(\langle\ell\rangle)$. Thus in this regime the wavelengths are selected by the granular dynamics scale that depends very little on transport while the associated propagation speeds depend very strongly on the transport regime. Note that these approximations are most accurate for exceedingly large $\langle\ell\rangle$ however Equations \eqref{eq: Omega_Dl(k) approx kl>>0} are still a reasonable approximation for Mars and Earth-like regimes (Figure \ref{fig4}) and support the general scaling of the model (Methods Figure \ref{fig-extended data:disp rel}). Interestingly, the term $\rho_{\Re}+\langle\mathcal{L}\rangle_{\Re}/\langle\cot\theta\rangle_n$ in $c(k)$ in eq. \eqref{eq: Omega_Dl(k) approx kl>>0} is positive for $\lambda\ll\langle\ell\rangle$ and negative for $\lambda\gtrsim\langle\ell\rangle$. Thus the approximation for $c$ in \eqref{eq: Omega_Dl(k) approx kl>>0} can approximate the ripple-antiripple scaling transition.



\subsection{Dispersion Relation in Terms of The Scale-Invariant Amplification Rate}\label{sec-supp: Dispersion Relation in Terms of The Scale-Free Amplification Rate}

Another more elegant way of seeing the wavelength selection mechanisms is by using the growth rate formalism in terms of $\mathcal{A}$ (Eq. \ref{eq-supp: Omega = int{A...}}). Like in the Methods, we define $\mathcal{A}-\delta \mathcal{A}=\mathcal{R} + i\mathcal{S}$ where $\mathcal{R}$ and $\mathcal{S}$ are the in-phase and out-of-phase contributions to the impact flux amplification, respectively, determined by hoplength changes $\mathcal{L}(k\ell)$. Here $\delta\mathcal{A}\simeq -ik\langle\varepsilon_{\Delta\ell}\rangle_n f_\uparrow e^{-ik\ell}$ is the small perturbation to the nearly scale-free forms of $\mathcal{R}(k\ell)$ and $\mathcal{S}(k\ell)$ (Fig. \ref{fig-extended data:A}). We note that in the Methods we ignore the small effects of $\delta\mathcal{A}$ for brevity and subtract $\delta\mathcal{A}\simeq -ik\langle\varepsilon_{\Delta\ell}\rangle_n f_\uparrow e^{-ik\ell}$ from direct DEM measurements of $\mathcal{A}$ in Methods Figure \ref{fig-extended data:A} to obtain the DEM values of $\mathcal{R}$ and $\mathcal{S}$ presented. Using these definitions we can express the dispersion relation as
\begin{eqnarray}\label{eq-supp: Omega(k) = int{R...+S}}
\frac{\sigma}{k\varphi_0} &=& \langle(1-n_\uparrow)\mathcal{R}\rangle - k\langle\Bar{\varepsilon} n_\uparrow\mathcal{S}\rangle \notag \\
\frac{c}{\varphi_0} &=& \langle(n_\uparrow-1)\mathcal{S}\rangle - k\langle\Bar{\varepsilon} n_\uparrow\mathcal{R}\rangle     
\end{eqnarray}
ignoring the contributions from $\delta\mathcal{A}$ for conceptual clarity. We saw in SI Section \ref{sec-supp: Modulation Rate Approximations} for large $\ell/\lambda\gtrsim 1$ around $k\sim k_*$ (small $\lambda/\langle\ell\rangle$), $\mathcal{A}\sim\mathcal{A}_*$ such that $\mathcal{R}_*\propto \sin k\ell$ and $\mathcal{S}_*\propto 1+\cos k\ell$ (Fig. \ref{fig-extended data:A}). Thus for nonlocal $\lambda/\langle\ell\rangle$, because most of the grains are leaving from the upwind face ($f_\uparrow\propto i$), $\mathcal{R}$ is peaked for small hoplengths around $\ell\sim \lambda/4$, acting like a `repton' amplification. Similarly, in the nonlocal regime the focusing effects ensure that $\mathcal{S}$ is peaked around $\ell\sim \lambda$, acting like a `salton' amplification. 

This way of stating the dispersion relation ultimately results in the same expressions derived above but now has a particularly intuitive meaning: The destabilizing mechanism now represents the average number of grains that remain on the crest after traveling from the upwind face $\langle(1-n_\uparrow)\mathcal{R}\rangle$. This effect is countered by the stabilization mechanism from the increase number of grains being eroded from the crest due to the down wind shift in the ejection flux from $k\langle\Bar{\varepsilon}n_\uparrow\mathcal{S}\rangle\sim k\langle\varepsilon_{\Delta\ell}\rangle_n \langle n_\uparrow\mathcal{S}\rangle$---this effect increases with the number of grains leaving the upwind face, $\langle n_\uparrow\mathcal{S}\rangle$.

Because the number of ejections/rebounds $n_\uparrow(\ell)$ is small for small hoplengths (Fig. \ref{fig3}), the speed contribution from $k\langle \Bar{\varepsilon}n_\uparrow\mathcal{R}\rangle$ is negligible thus leaving the propagation speed to be determined by the number of grains eroded from or deposited on the upwind face $\langle(n_\uparrow-1)\mathcal{S}\rangle$ which is positive in the nonlocal regime where $f_\uparrow\propto i$ because grains are being distributed from the upwind face over to the back of the ripple. However, if the wavelength is selected in the local regime (small $\langle\ell\rangle$) then the speeds can be negative.

With these considerations we arrive at the approximation
\begin{eqnarray}\label{eq: Omega(k) = <R> -<S>}
    \frac{\sigma}{k\varphi_0} &\sim& \langle(1-n_\uparrow)\mathcal{R}\rangle - k\langle\varepsilon_{\Delta\ell}\rangle_n\langle n_\uparrow\mathcal{S}\rangle  \notag\\
     \frac{c}{\varphi_0} &\sim& \langle(n_\uparrow-1)\mathcal{S}\rangle
\end{eqnarray}
which reduces to the main text finding that 
\[\frac{\lambda}{\langle\varepsilon_{\Delta\ell}\rangle_n}\propto \frac{\langle n_\uparrow\mathcal{S}\rangle}{\langle(1-n_\uparrow)\mathcal{R}\rangle}\]
with a proportionality factor $a(\langle\ell\rangle)$ associated with the contribution from $\delta\mathcal{A}$.

\subsection{The Transport Regime Sets the Propagation Speed}\label{sec-supp: speed selection}
As we have seen above the fastest growing wavelength can be written generally as $\lambda/\langle \varepsilon_{\Delta\ell}\rangle_n \simeq f(\langle\ell\rangle/d)$. The factor $f(\langle\ell\rangle/d)$ is an almost unique function of $\langle\ell\rangle/d$ and so is $\langle \varepsilon_{\Delta\ell}\rangle_n/d$. Thus the dispersion relation can be expressed as a function of two groupings, $k\langle \ell \rangle$ and $\langle\ell\rangle/d$:
\[\frac{\Omega}{k\varphi_0} = G(k\langle\ell\rangle,\langle\ell\rangle/d).\]
The fastest growing wavelength follows from the condition $\partial_k\sigma=0$ which renders $k\langle\ell\rangle=g(\langle\ell\rangle/d)$ and thus the speed is found to be a unique function of $\langle\ell\rangle/d$:
\[\frac{c}{\varphi_0}=F(\langle\ell\rangle/d)\]
recovering the previous approximation that while $\lambda$ scales with a granular length $\langle \varepsilon_{\Delta\ell}\rangle_n\propto d$, the speed $c$ scales with the transport regime, $\varphi_0$ and $\langle\ell\rangle/d $ (Fig. \ref{fig-extended data:disp rel}).


\end{document}